\documentclass[iopscience,apj]{emulateapj}
\usepackage{graphicx,epsfig}
\usepackage{amsmath, amsthm, amssymb}
\usepackage{times}
\usepackage{soul}
\usepackage[plainpages=false, colorlinks=true, anchorcolor=blue, linkcolor=blue,
 citecolor=blue, bookmarks=false]{hyperref}

\usepackage{graphicx,amsmath}

\newcommand{\jump}[1]{\Delta_s \left( #1 \right)}
\newcommand\kms{{\rm\, km\, s^{-1}}}

\newcommand\kpc{{\rm\, kpc}}
\newcommand\cs{c_s}
\newcommand\cA{\mathcal{A}}
\newcommand\F{\mathcal{F}}

\newcommand\omgSP{\tilde \omega_{D, \rm sp}}
\newcommand\tky{\tilde k_y}
\newcommand{\omgRI}[1]{{\rm Re}(\omega_{#1})}
\newcommand{\omgR}{{\rm Re}(\omega)}
\newcommand{\omgI}{{\rm Im}(\omega)}
\newcommand\uT{u_{T0}}
\newcommand\vT{v_{T0}}

\newcommand\spost{{s+}}
\newcommand\spre{{s-}}
\newcommand\DX{\delta \eta}
\newcommand\xSO{x_{\rm sp}}
\newcommand\xSH{x_{\rm sh}}
\newcommand\kx{k_x}
\newcommand\kxa{k_{x,a}}
\newcommand\kxv{k_{x,v}}
\newcommand\ky{k_y}
\newcommand\freq{{\rm\, km\, s^{-1}\, kpc^{-1}}}
\newcommand\tgrow{t_{\rm grow}}
\newcommand\simgt{\lower.5ex\hbox{$\; \buildrel > \over \sim \;$}}
\newcommand\simlt{\lower.5ex\hbox{$\; \buildrel < \over \sim \;$}}

\slugcomment{\today}

\begin{document}

\title{Nature of the Wiggle Instability of Galactic Spiral Shocks}

\author{Woong-Tae Kim, Yonghwi Kim, \& Jeong-Gyu Kim}

\affil{Center for the Exploration of the Origin of the Universe (CEOU),
Astronomy Program, Department of Physics \& Astronomy,\\
Seoul National University, Seoul 151-742, Republic of Korea}

\email{wkim@astro.snu.ac.kr, kimyh@astro.snu.ac.kr,
jgkim@astro.snu.ac.kr}

\begin{abstract}
Gas in disk galaxies interacts nonlinearly with a underlying stellar
spiral potential to form galactic spiral shocks. While numerical
simulations typically show that spiral shocks are unstable to wiggle
instability (WI) even in the absence of magnetic fields and
self-gravity, its physical nature has remained uncertain. To clarify
the mechanism behind the WI, we conduct a normal-mode linear
stability analysis as well as nonlinear simulations assuming that
the disk is isothermal and infinitesimally thin. We find that the WI
is physical, originating from the generation of potential vorticity
at a deformed shock front, rather than Kelvin-Helmholtz
instabilities as previously thought. Since gas in galaxy rotation
periodically passes through the shocks multiple times, the potential
vorticity can accumulate successively, setting up a normal mode that
grows exponentially with time. Eigenfunctions of the WI decay
exponentially downstream from the shock front. Both shock
compression of acoustic waves and a discontinuity of shear across
the shock stabilize the WI. The wavelength and growth time of the WI
depend on the arm strength quite sensitively. When the stellar-arm
forcing is moderate at 5\%, the wavelength of the most unstable mode
is about 0.07 times the arm-to-arm spacing, with the growth rate
comparable to the orbital angular frequency, which is found to be in
good agreement with the results of numerical simulations.
\end{abstract}

\keywords{galaxies: ISM -- galaxies: kinematics and dynamics --
galaxies: spiral -- galaxies: structure   -- hydrodynamics --
instabilities --- ISM: general -- shock waves -- stars: formation}

\section{Introduction}

Spiral arms are the most prominent structures in disk galaxies,
playing a vital role in their secular evolution (e.g.,
\citealt{but96,kor04,but13,sel14}).  They possess secondary
structures such as young stellar complexes and \ion{H}{2} regions
distributed in a ``beads on a string" fashion along them (e.g.,
\citealt{baa63,elm83,elm06,she07}) as well as giant molecular clouds
in which new star formation takes place (e.g.,
\citealt{vog88,ran93,sak99,kod09,sch13,mei13}). Another secondary
feature includes gaseous feathers, referring to filamentary
structures that protrude almost perpendicularly from the arms and
are swept into a trailing configuration in the interarm regions,
seen in optical or infrared images of nearby spiral galaxies (e.g.,
\citealt{scorec01,sco01,ken04,wil04,lav06,cor08,sil12,sch13}). These
are in close geometrical association with narrow dust lanes that
represent shocked interstellar gas due to its gravitational
interaction with the stellar spiral arms.  This strongly suggests
that the shock compression of gas in galaxy rotation may trigger
formation of the secondary structures and ensuing star formation in
galactic disks.

One of the unsolved problems regarding galactic spiral shocks is
what mechanism is responsible for the secondary structure formation
after the shock compression. There have been a number of studies on
this issue (e.g.,
\citealt{bal85,bal88,elm94,kim02,kim06,wad04,she06,she08,dob06,dob07,
lee12}), but they differ in the relative importance of gas
self-gravity, magnetic fields, and other hydrodynamic processes. For
example, \citet{bal88} used a Lagrangian linear stability analysis
of postshock flows and showed that self-gravity allows hydrodynamic
disturbances to grow transiently via swing amplifier. \citet{elm94}
found that an inclusion of azimuthal magnetic fields gets rid of a
stabilizing effect of epicycle motions, making self-gravity more
powerful in gathering the gas.

\citet{kim02,kim06} ran direct numerical simulations using local
shearing-box models and found that magnetized disturbances indeed
grow much faster than the case of pure swing amplification, rapidly
forming perpendicular structures that resemble observed feathers.
They further showed that these feather-like structures experience
fragmentation at the nonlinear stage and form gravitationally bound
clouds.  These results based on local models were shown valid also
in global simulations of \citet{she06}. More recently, \citet{lee12}
performed an Eulerian linear stability analysis of the shearing-box
models considered by \citet{kim02,kim06}, and found that feathers
represent parasitic instabilities intrinsic to a self-gravitating,
magnetized spiral shock.  Interestingly, these feather-forming
instabilities are referred to differently as azimuthal instability,
magneto-Jeans instability, and feathering instability in
\citet{elm94}, \citet{kim02}, and \citet{lee12}, respectively.

On the other hand, numerical studies of \citet{joh86},
\citet{wad04}, \citet{dob_etal06}, and \citet{dob06,dob07} have
shown that self-gravity is not prerequisite to the formation of
secondary structures.  In particular, \citet{wad04} ran global,
non-self-gravitating simulations of galactic disks with no magnetic
field, and found that spiral shocks are unstable to wiggling
perturbations and form dense clumps in the shock-compressed layer,
provided that the shocks are strong.  They termed this clump-forming
hydrodynamic instability the wiggle instability (WI), and suggested
it as a feather formation mechanism. \citet{dob06} also observed
formation of clumps along the arms and feathers projecting into the
interarm regions in their smoothed particle hydrodynamics (SPH)
simulations, which they attributed to orbit crowding of particles
with non-uniform density that change their angular momenta in the
shock. In the models of \citet{dob06}, it is necessary for the gas
to be cold to grow into interarm features. In a more recent
high-resolution numerical study with self-gravity and star-formation
feedback included, \citet{ren13} found that spiral shocks produce
regularly-spaced, star-forming clumps due to a strong velocity
gradient near the arms.

The WI of large-scale galactic shocks appears ubiquitous in
hydrodynamic simulations of disk galaxies with non-axisymmetric
patterns as long as resolution is large enough to resolve it. For
instance, recent grid-based simulations of \citet{kim14} for radial
mass drift by spiral shocks found a strong development of the WI at
the shock fronts that grows faster at smaller radii except near the
corotation resonance. This suggests that the WI directly involves
shock fronts and grows within an orbital time scale (see also
\citealt{she06}). In addition to spiral shocks, dust-lane shocks
surrounding a nuclear ring in barred galaxies are found to suffer
from WI to form dense clumps along them (e.g.,
\citealt{kim12a,kim12b,kim_sto12,seo13}).  The WI seems to be
suppressed by magnetic fields pervasive in the interstellar medium
\citep{she06,dob_price08} and by shock flapping motions that
naturally occur in a vertically stratified disk
\citep{kim06,kimcg06,kimcg10}.

Despite these numerical efforts, however, the physical nature of the
WI has remained elusive so far. Based on a Richardson-number
criterion, \citet{wad04} proposed that the WI originates in
Kelvin-Helmholtz instabilities (KHI) occurring in a shear layer
behind the shock.  \citet{ren13} also noted that the morphologies of
clumps formed in their simulations are similar to the patterns in
the KHI. However, the numerical simulations mentioned above show
that the instability grows from the shock front distortion itself,
and an isolated shock front has been known unconditionally stable to
distortional perturbations (see below). In addition, as
\citet{wad04} noted, the Richardson-number criterion is only a
\emph{necessary} condition for \emph{stability} \citep{cha61}, so
that one should be cautious when applying it as an instability
criterion.  Moreover, \citet{dwa96} showed that the postshock flow
of a spiral shock is linearly stable to the KHI. \citet{wan10} also
argued that shear in the postshock flow can be removed by choosing a
frame moving at the tangential velocity at the shock front, implying
that the postshock layer is stable to the KHI.  On the other hand,
\citet{kim12a} found that the WI of dust-lane shocks in barred
galaxies is deeply related to the growth of potential vorticity (PV)
from curved spiral shocks. In a quite different perspective,
\citet{han12} raised a possibility that the WI may be due entirely
to numerical artifacts arising from the inability to resolve a shock
inclined to numerical grids.

In this paper, we perform a linear stability analysis of galactic
spiral shocks, aiming to clarify the physical nature of the WI. We
adopt a local shearing-box model of an infinitesimally-thin galactic
gaseous disk, and assume that the gas is isothermal, unmagnetized,
and non-self-gravitating.  This simple disk model is of course
unrealistic in the sense that it cannot handle important physics
related to the multi-phase, turbulent interstellar medium (e.g.,
\citealt{fie69,mck77,wol03,elm04}), star formation, feedback, etc.,
and is unable to capture the curvature effect of large-scale shocks.
Nevertheless, it incorporates all necessary ingredients to explore
spiral shocks (e.g., \citealt{haw95,kim02}), and allows to study WI
at a fixed angular frequency of galaxy rotation. Technically, we
follow the Eulerian linear stability analysis presented by
\citet{lee12}, but neglect the effects of magnetic fields and
self-gravity in the present work.

Stability of an isolated, planar, two-dimensional shock front in an
inviscid medium has been studied intensively in the fluid dynamics
community (e.g., \citealt{dya54,fre57,van75,swa75}; see also
\citealt{lan87}). Here, the term ``isolated" indicates a situation
where a shock is located far away from its driving source and the
upstream supersonic flow is completely unperturbed. The general
result is that sinusoidal wiggling perturbations to an otherwise
planar shock decay asymptotically with time as an inverse power law,
and such shocks are unconditionally stable if the gas follows an
isothermal or an ideal-gas equation of state (e.g.,
\citealt{rob00,bat07})\footnote{Under an arbitrary equation of
state, isolated planar shocks can be unstable if certain conditions
are met (e.g., \citealt{swa75,bat07}), the discussion of which is
beyond the scope of the present paper.}.  This suggests that the WI
of an isothermal spiral shock, if it is physical, must involve
perturbations in the preshock regions. We shall show below that it
makes use of PV generated from a perturbed shock front. Spiral
shocks cannot be treated isolated since gas in galaxies crosses them
multiple times in the course of galaxy rotation. If perturbations
remain coherent before and after the shock fronts, PV can grow
continuously through successive passages of spiral shocks, leading
to the WI. In addition to galactic disks, the PV generation by
curved shocks also actively engages  in the dynamics of
protoplanetary disks. The PV accumulation by repeated passages of
shocks produced by an embedded planet is known to give rise to a
secondary instability near the corotation resonance, changing the
gravitational torque on and thus the migration time scale of the
planet (e.g., \citealt{bal01,kol03,li05,dev07,lin12}).

The remainder of this paper is organized as follows. In Section
\ref{s:beqn}, we describe the basic equations we solve and specify
the parameters we adopt.  In Section \ref{s:back}, we obtain the
steady equilibrium solutions of spiral shocks that we use as a
background state of the WI. In Section \ref{s:method}, we present
the formulation of our normal-mode stability analysis, the shock
jump conditions, and the spatial behavior of PV that perturbations
should obey, and the computation method to find eigenvalues.   The
resulting dispersion relations for one- and two-dimensional modes
together with physical interpretation in terms of PV are presented
in Section \ref{s:disp}.  In Section \ref{s:num}, we run direct
numerical simulations of the WI, and compare the results with those
of the linear stability analysis.  In Section \ref{s:sum}, we
conclude with a summary and discussion of our results in comparison
with the previous studies.

\section{Basic Equations}\label{s:beqn}

We consider an infinitesimally-thin, non-self-gravitating galactic
gaseous disk with no magnetic field, and study its responses to an
imposed stellar spiral-arm potential.  The disk is rotating at
angular frequency $\Omega$ at the galactocentric radius $R$. We
assume that the gas is isothermal with a sound speed of $\cs$.  The
arms rotate rigidly about the galaxy center at a fixed pattern speed
$\Omega_p$.

For problems involving spiral arms, it is advantageous to employ a
local Cartesian frame $(x, y)$ corotating with the arms lying at
$R$, introduced by \citet{rob69}. In this frame,  the two orthogonal
$x$- and $y$-axes correspond to the directions perpendicular and
parallel to the local spiral arm, respectively (see also
\citealt{rob70,shu73,bal88,kim02,lee12}). We make a local
approximation ($|x|, |y| \ll R$) and assume that the arms are
tightly wound with a pitch angle $i$.  In the absence of the
spiral-arm potential, the gas has the uniform surface density
$\Sigma_c$ and the velocity $\mathbf{v}_c \equiv (u_c, v_c)$, where
\begin{equation}\label{e:vc}
u_c  = R(\Omega-\Omega_p)\sin i, \;\;\;\text{and}\;\;\; v_c =
R(\Omega-\Omega_p) - q\Omega x,
\end{equation}
arising from galaxy rotation. Here, $q \equiv -d\ln\Omega /d\ln R$
measures local shear rate and is equal to unity for flat rotation.
The basic equations of ideal hydrodynamics expanded in this local
frame read
\begin{equation}\label{e:con}
  \frac{\partial\Sigma}{\partial t} +
  \nabla\cdot(\Sigma \mathbf{v}_T) = 0,
\end{equation}
\begin{equation}\label{e:mom}
  \frac{\partial\mathbf{v}}{\partial t} +
        \mathbf{v}_T\cdot\nabla\mathbf{v}
     =  -{\cs^2}\nabla \ln\Sigma + q\Omega u \mathbf{\hat{y}}
       - 2\mathbf{\Omega}\times\mathbf{v}
       - \nabla \Phi_s,
\end{equation}
where $\Sigma$ is the gas surface density, $\mathbf{v}\equiv (u, v)$
is the velocity induced by the arms, $\mathbf{v}_T = \mathbf{v} +
\mathbf{v}_c$ is the total velocity in the local frame, and $\Phi_s$
is the stellar spiral-arm potential (e.g., \citealt{kim02}).

For an $m$-armed spiral, the arm-to-arm distance along the
$x$-direction is $L =2\pi R\sin i /m$.  To be consistent with the
local approximation, gas flows should be periodic in the
$x$-direction, with period $L$ in length.  For the external spiral
potential, therefore, we take a simple form
\begin{equation}\label{e:extP}
\Phi_s = \Phi_0 \cos \left(\frac{2\pi x}{L}\right),
\end{equation}
with amplitude $\Phi_0 ~(>0)$, which is  a local analog of a
logarithmic potential considered by \citet{rob69} and \citet{shu73}.
We confine to the domain with $0\leq x \leq L$, so that the
potential minimum occurs at the center of the domain (i.e.,
$x=L/2$). We parameterize $\Phi_0$ using a dimensionless parameter
\begin{equation}\label{e:F}
 \F\equiv \frac{m}{\sin i} \left(\frac{\Phi_0}{R^2\Omega^2}\right),
\end{equation}
which measures the maximum force due to the spiral arm relative to
the centrifugal force of galaxy rotation (e.g., \citealt{rob69}).

In two-dimensional gas flows, conservation of both angular momentum
and mass leads to conservation of PV defined by
\begin{equation}\label{e:pvdef}
\boldsymbol{\xi} \equiv  \frac{\nabla\times
\mathbf{v}_T+2{\boldsymbol\Omega}}{\Sigma}.
\end{equation}
Using equations (\ref{e:con}) and (\ref{e:mom}), one can directly
show that
\begin{equation}\label{e:pv}
\left(\frac{\partial}{\partial t}
    + \mathbf{v}_T \cdot\nabla
\right) \boldsymbol{\xi} = 0,
\end{equation}
indicating that $\boldsymbol{\xi}$ remains unchanged along a given
streamline (e.g., \citealt{hun64,gam96}), provided that it does not
intersect curved discontinuities such as shocks or contact
discontinuities.  We will show in Section \ref{s:disp} that a
deformed shock front can serve as a source of PV, which in turn
renders the shock prone to the WI.

Equations (\ref{e:con})--(\ref{e:F}) are completely specified by six
dimensionless parameters: $q$, $m$, $\sin i$, $\Omega_p/\Omega$,
$\F$, and $\cs/(R\Omega)$.  For our numerical examples presented
below, we take $q=1$, $m=2$, $\sin i=0.1$, $\Omega_p/\Omega=0.5$,
$\F=3$--$10\%$, and $\cs/(R\Omega)=0.027
(\cs/7\kms)(\Omega/26\freq)^{-1}(R/10\kpc)^{-1}$, which represents
conditions in normal disk galaxies fairly well.

We remark a few limitations of our local models.  First, they
neglect terms arising from curvature effects in the coordinates that
may be important for forming large-scale shocks associated with arms
with a large pitch angle. Second, the local approximation with $\sin
i\ll 1$ tends to make $\F$ smaller than realistic values in disk
galaxies with not-so-tightly-wound arms (see Eq.\
[\ref{e:F}])\footnote{We found that $\F=5\%$ when $\sin i=0.1$
produces equilibrium spiral shocks that are equivalent to
$\F=10-12\%$ when $\sin i=0.34$ (or $i=20^\circ$, corresponding to
the arms of M51).}. Third, limited to the galactic midplane, our
two-dimensional models are unable to capture non-planar dynamics,
such as shock flapping motions \citep{kim06}, involving the vertical
dimension.  Nevertheless, our models under the shearing box
approximation naturally incorporate large scale shear, and can thus
well describe periodic gas flows as well as epicycle motions, which
are the essential ingredients of galactic spiral shocks. Our local
models are in fact ideal to identify the physical mechanism behind
the WI qualitatively, as we will present in Section \ref{s:disp}.

\section{Background State}\label{s:back}

As a first step, we seek for steady-state solutions, $\Sigma_0(x)$,
$u_0(x)$, $v_0(x)$, of equations (\ref{e:con}) and (\ref{e:mom}).
Here and hereafter, we use the subscript ``0'' to indicate the
time-independent shock solutions. Such solutions were obtained by
previous studies (e.g., \citealt{rob69,shu72,shu73,kim02,git04}). We
revisit this issue here in order to obtain a background state of the
WI.

The steady solutions of spiral shocks satisfy
\begin{equation}\label{e:con0}
\Sigma_0 \uT =  \Sigma_{c} u_c = \text{constant},
\end{equation}
\begin{equation}\label{e:momx0}
\uT \frac{du_0}{dx} = -\frac{c_s^2}{\Sigma_0}\frac{d\Sigma_0}{dx} +
2 \Omega v_ 0 - \frac{d\Phi_s}{dx},
\end{equation}
and
\begin{equation}\label{e:momy0}
\uT\frac{dv_0}{dx} = -\frac{\kappa^2}{2\Omega} u_0,
\end{equation}
where $\kappa^2 = R^{-3}d(\Omega^2 R^4)/dR=(4-2q)\Omega^2$ is the
square of the epicycle frequency. Equations (\ref{e:con0}) and
(\ref{e:momx0}) are combined to give
\begin{equation}\label{e:momx01}
\left(\uT - \frac{\cs^2}{\uT} \right)\frac{d\uT}{dx} = 2\Omega v_0 +
R\Omega^2 \F \sin\left(\frac{2\pi x}{L}\right).
\end{equation}
Equations (\ref{e:momy0}) and (\ref{e:momx01}) can be solved
numerically  over $0\leq x \leq L$ subject to the periodic boundary
conditions at $x=0$ and $L$.

Since
\begin{equation}\label{e:aux1}
  \frac{d\vT}{dx} =
  \frac{\kappa^2}{2\Omega} \frac{u_c}{\uT} -2 \Omega,
\end{equation}
it follows that
\begin{equation}\label{e:pv0}
  \xi_0 = \frac{|\nabla\times \mathbf{v}_{T0} +
  2\mathbf{\Omega}|}{\Sigma_0} =
 \frac{\kappa^2}{2\Omega\Sigma_c},
\end{equation}
showing that the PV of the steady spiral shocks is constant
everywhere.

\subsection{Expansion near the Sonic Point}\label{s:sonic0}

\citet{shu73} showed that even a very weak spiral forcing ($\F >
0.9\%$ for their model parameters) results in shocks in the gas
flow. In order to meet the periodic boundary conditions, spiral
shocks should involve a sonic point where $\uT=\cs$, through which a
subsonic gas accelerates to supersonic speeds. Let $\xSO$ denote the
location of the sonic point. The right-hand side of equation
(\ref{e:momx01}) should vanish at $x=\xSO$ for a transonic solution
to exist.  Let us expand $\uT$ and $v_0$ around the sonic point as
\begin{mathletters}\label{e:so0}
\begin{eqnarray}
\uT/(R\Omega) &=& a + \alpha_1 \DX + \alpha_2 \DX^2 + \mathcal O
(\DX^3), \label{e:sou0}\\
v_0/(R\Omega) &=&  \beta_0 + \beta_1 \DX  + \beta_2 \DX^2 + \mathcal
O(\DX^3), \label{e:sov0}
\end{eqnarray}
\end{mathletters}
where $a\equiv \cs/(R\Omega)$, $\DX \equiv (x -\xSO)/R$, and the
coefficients $\alpha_{1,2}$ and $\beta_{0,1,2}$ are to be
determined. Here, we keep up to second-order terms in the series
expansion of the velocities since they are needed in the expansion
of the perturbation variables in Section \ref{s:sonic1}.

Plugging equation (\ref{e:so0}) into equations (\ref{e:momy0}) and
(\ref{e:momx01}), we find that
\begin{equation}\label{e:bet0}
\beta_0 = -\frac{\F}{2} \sin \left( \frac{2\pi \xSO}{L}\right),
\end{equation}
\begin{equation}\label{e:alp1}
\alpha_1 = \left[\beta_1 + \frac{\F}{\sin i} \cos\left( \frac{2\pi
\xSO}{L}\right)\right]^{1/2},
\end{equation}
\begin{equation}\label{e:bet1}
\beta_1 = \frac{u_c - \cs}{\cs},
\end{equation}
\begin{equation}\label{e:alp2}
\alpha_2 = \frac{\alpha_1^2}{6a} - \frac{u_c/(R\Omega)}{6a^2} +
\frac{2\beta_0}{3\alpha_1\sin^2i},
\end{equation}
and
\begin{equation}\label{e:bet2}
\beta_2 = -\frac{1+\beta_1}{2a}\alpha_1.
\end{equation}
Note that we take a positive sign for $\alpha_1$ in equation
(\ref{e:alp1}) since $\uT$ should increase across the sonic point
for a transonic solution.

\subsection{Jump Conditions}

In addition to the periodic boundary conditions at $x/L=0$ and $1$,
the spiral-shock solutions should satisfy the following jump
conditions at the shock front, $x=\xSH$:
\begin{mathletters}\label{e:sj}
\begin{eqnarray}
\jump{ \uT \Sigma_0            } & = & 0, \label{e:s1} \\
\jump{ (\cs^2 + \uT^2)\Sigma_0 } & = & 0, \label{e:s2} \\
\jump{ v_0                     } & = & 0, \label{e:s3}
\end{eqnarray}
\end{mathletters}
where  $\jump{f} \equiv f^\spost - f^\spre$, with the superscripts
``$\spost$'' and ``$\spre$'' indicating the quantities evaluated at
the immediate behind ($x=\xSH + 0$) and ahead ($x=\xSH -0$) of the
shock front, respectively.

Equation (\ref{e:s1}) is automatically satisfied from equation
(\ref{e:con0}).  Equation (\ref{e:s2}) is equivalent to
\begin{equation}\label{e:bd1}
\uT^\spost \uT^\spre = \cs^2,
\end{equation}
a jump condition for the perpendicular velocity in an isothermal
shock, while equation (\ref{e:s3}) states that the parallel
component of the velocity should be continuous across the shock.

\begin{deluxetable}{cccc}
\tablecaption{Properties of Equilibrium Spiral Shocks\label{t:1d}}
\tablewidth{0pt} %
\tablehead{ \colhead{$\F$} & \colhead{$\xSO/L$}   &
\colhead{$\xSH/L$} & \colhead{$\mu$}} \startdata
0.03   & 0.458 & 0.402 & 5.67 \\
0.05   & 0.507 & 0.431 & 11.6 \\
0.10   & 0.615 & 0.495 & 34.5 \\
0.20   & 0.699 & 0.560 & 97.3
\enddata
\tablecomments{For the arm and galaxy parameters of $q = 1$, $m =
2$, $\sin i = 0.1$, $\Omega_p/\Omega=0.5$, and
$\cs/(R\Omega)=0.027$.}
\end{deluxetable}

\subsection{Equilibrium Shock Structure}

Since we do not know the locations of the sonic point and the shock
a priori, we first choose $\xSO$ arbitrarily for given $\F$ and then
integrate equations (\ref{e:momy0}) and (\ref{e:momx01}) starting
from $\xSO$ in both forward and backward directions, noting that
$u_0$ and $v_0$ are periodic at $x/L=0$ and $1$. We determine $\xSH$
from equation (\ref{e:s3}), and check the jump condition for the
perpendicular velocity at that location. If equation (\ref{e:bd1})
is not satisfied, we return to the first step, and repeat the
calculation by changing $\xSO$ iteratively until all the jump
conditions are met within tolerance (typically $\sim 10^{-5}$).
Table \ref{t:1d} lists the values of $\xSO$, $\xSH$, and the density
jump defined by
\begin{equation}\label{e:mu}
\mu = \Sigma_0^\spost/\Sigma_0^\spre,
\end{equation}
for a few selected values of $\F$.

\begin{figure}
\epsscale{1.1}\plotone{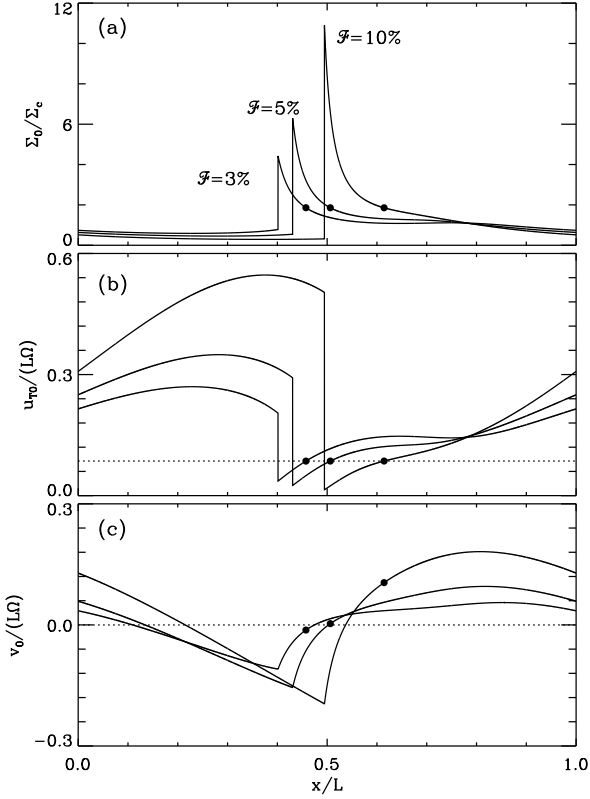} \caption{One-dimensional
steady-state shock profiles for $\F=3$, 5, and 10\%. In (b), the
horizontal dotted line indicates the sound speed. Each dot marks the
sonic point for the corresponding $\F$.\label{f:1d}}
\end{figure}

Figure \ref{f:1d} illustrates equilibrium structures of
one-dimensional spiral shocks for $\F=3$, 5, and 10\%.  The gas is
flowing from left to right.  In each panel, dots mark the sonic
points. The dotted line in Figure \ref{f:1d}(b) indicates the sound
speed $\cs$. When a gas element hits the shock front, it is
compressed to suffer a density jump, which occurs at the expense of
a decrease in $\uT$. In order for the flow to be periodic, the gas
should be accelerated downstream and pass through the sonic point,
increasing $\uT$.  The constraint of the potential vorticity
conservation requires the velocity parallel to the shock to increase
as $\Omega^{-1}d\vT/dx= (2-q)(\Sigma_0/\Sigma_c)-2$ after the shock,
making shear reversed wherever $\Sigma_0/\Sigma_c> 2/(2-q)$
\citep{bal85,kim01,kim02}. Since $\Sigma_0(\xSO)/\Sigma_c = u_c/\cs
= 1.86$ for our adopted set of parameters, the sonic point lies just
outside the region of shear reversal.

To further understand the behaviors of the steady solutions near the
sonic point, it is useful to consider the problem in analogy with a
flow through the De Laval nozzle of a jet engine, for which any
smooth sonic transition occurs only when the cross-sectional area
$\cA(x)$ achieves a local minimum at the sonic point (see, e.g.,
\citealt{shu92}). By comparing equation (6.28) of \citet{shu92} with
equation (\ref{e:momx01}), one can see that two processes (epicycle
shaking and gravitational focusing of the arm) shape $\cA(x)$ for
galactic gas flows. The two conditions ($d\cA/dx=0$ and
$d^2\cA/dx^2>0$) for the minimum cross-sectional area at the sonic
point give $\beta_0$ in equation (\ref{e:bet0}) and require the
positivity of the terms inside the square brackets in equation
(\ref{e:alp1}).  Obviously, the gravitational focusing term
increases (or decreases) $\cA$ with $x$ where $\xSO < L/2$ (or where
$\xSO > L/2$), which in turn requires the Coriolis term should
decrease (or increase) $\cA$.  Thus, $v_0(\xSO) <0$ when $\xSO<L/2$,
which happens when the spiral forcing is weak with $\F<5\%$, while
$v_0(\xSO) > 0$ when $\xSO> L/2$, as Figure \ref{f:1d} shows. In the
former (latter) case, the effect of the gravitating focusing
relative the Coriolis term becomes smaller (larger) as $\F$
increases, which tends to shift the sonic point toward the
downstream side.

\section{Normal-mode Linear Stability Analysis}\label{s:method}

\subsection{Perturbation Equation}

We now focus on the main theme, the normal-mode linear stability
analysis of one-dimensional steady solutions of spiral shocks found
in the preceding section.  Upon top of the equilibrium profiles
$\Sigma_0$, $u_0$, and $v_0$, we impose small-amplitude
perturbations $\Sigma_1$, $u_1$, and $v_1$. Assuming that the
perturbed quantities are much smaller than the background values, we
linearize equations (\ref{e:con}) and (\ref{e:mom}) to obtain
\begin{align}\label{e:con1}
\frac{\partial}{\partial t}\left(\frac{\Sigma_1}{\Sigma_0}\right) & +
\left(\uT\frac{\partial}{\partial x} + \vT\frac{\partial}{\partial
y}\right)\left(\frac{\Sigma_1}{\Sigma_0}\right) \nonumber \\ & + \frac{\partial
u_1}{\partial x} + \frac{d\ln\Sigma_0}{dx} u_1 + \frac{\partial
v_1}{\partial y}=0,
\end{align}
\begin{align}\label{e:mx1}
\frac{\partial u_1}{\partial t} & + \left(\uT \frac{\partial
u_1}{\partial x} + \vT\frac{\partial u_1}{\partial y}\right) \nonumber \\ & +
\frac{du_0}{dx} u_1 + \cs^2 \frac{\partial}{\partial
 x}\left(\frac{\Sigma_1}{\Sigma_0}\right) - 2 \Omega v_1=0,
\end{align}
\begin{align}\label{e:my1}
\frac{\partial v_1}{\partial t} & + \left(\uT \frac{\partial
v_1}{\partial x} + \vT\frac{\partial v_1}{\partial y}\right) \nonumber \\ & + \cs^2
\frac{\partial}{\partial
 y}\left(\frac{\Sigma_1}{\Sigma_0}\right) + \left(\frac{\kappa^2}{2\Omega}\right)\frac{u_c}{\uT}
 u_1=0.
\end{align}

Since the coefficients in equations (\ref{e:con1})--(\ref{e:my1})
depend only on $x$ and are independent of $t$ and $y$, we consider
perturbations of the form
\begin{equation}\label{e:ptb}
\left(
 \begin{array}{c}
 \Sigma_1/\Sigma_0 \\
 u_1 \\
 v_1
\end{array}\right) =
\left(
 \begin{array}{c}
 S_1(x)\\
 U_1(x)\\
 V_1(x)
\end{array}\right)
 \exp(-i\omega t + ik_yy),
\end{equation}
where $\omega$ and $k_y$ denote the perturbation frequency and
wavenumber in the $y$-direction, respectively. Equations
(\ref{e:con1})--(\ref{e:my1}) then reduce to
\begin{align}\label{e:ds1}
(\uT^2 &-\cs^2)\frac{dS_1}{dx} =
 i\uT\omega_D S_1  \nonumber \\ & +
 \left(2\frac{du_0}{dx} - i\omega_D \right)U_1 - (ik_y\uT + 2\Omega)
 V_1,
\end{align}
\begin{align}\label{e:du1}
& (\uT^2-\cs^2) \frac{dU_1}{dx} =
 - i \cs ^2\omega_D S_1 
 \nonumber \\ & - \left[
 \frac{\cs^2}{\uT}\frac{du_0}{dx} -
 \uT\left(i\omega_D - \frac{du_0}{dx}\right)\right]U_1
 + (ik_y \cs^2 + 2\uT\Omega) V_1,
\end{align}
\begin{equation}\label{e:dv1}
 \uT\frac{dV_1}{dx} = -ik_y \cs^2 S_1 - \frac{\kappa^2}{2\Omega}\frac{u_c}{\uT} U_1 +
 i\omega_D V_1,
\end{equation}
where
\begin{equation}\label{e:dOm}
 \omega_D (x) = \omega - k_y \vT,
\end{equation}
is the Doppler-shifted frequency. These are our perturbation
equations that can be integrated over $x$ as an eigenvalue problem
to find eigenvalue $\omega$, subject to the proper boundary
conditions.  We take a convention that $\omega$ is complex, while
$\ky$ is a pure real number.

\subsection{Perturbed Potential Vorticity}

By applying perturbations to equation (\ref{e:pvdef}), we obtain the
perturbed PV
\begin{equation}\label{e:pvv1}
\xi_1 = \frac{|\nabla \times \mathbf{v_1}|}{\Sigma_0}  - \xi_0
\frac{\Sigma_1}{\Sigma_0}.
\end{equation}
Analogous to equation (\ref{e:ptb}), we define the amplitude $\Xi_1$
of $\xi_1$ as $\Xi_1(x)\equiv \xi_1 (x,y,t) e^{i\omega t - i\ky y}$.
In terms of the perturbation variables, equation (\ref{e:pv}) then
becomes
\begin{equation}\label{e:pvcon1}
\left(-i\omega_D + \uT\frac{d}{dx}\right) \Xi_1 = 0,
\end{equation}
where
\begin{equation}\label{e:pv1}
\Xi_1 = \frac{1}{\Sigma_0}\left( \frac{dV_1}{dx} -ik_y U_1 \right) -
\xi_0 S_1.
\end{equation}

Equation (\ref{e:pvcon1}) is integrated to yield
\begin{equation}\label{e:pvint1}
  \frac{\Xi_1(x)}{\Xi_1^\spost} = e^{-\tau\omgI}
  \exp\left(i \int_{\xSH}^x \kxv (x) dx \right),
\end{equation}
where $\tau$ is the Lagrangian time
\begin{equation}\label{e:tau}
\tau \equiv \int_{\xSH}^x \frac{dx}{\uT},
\end{equation}
starting from the shock front, and
\begin{equation}\label{e:pvint2}
  \kxv \equiv  \frac{\omgR -\vT\ky}{\uT},
\end{equation}
represents the local $x$-wavenumber of the perturbed PV (or, more
generally, the entropy-vortex mode).

It is straightforward to show that the trajectory of constant phase
of the perturbed PV on the $x$-$y$ plane is described by $dy/dx =
-\kxv/\ky=\mathcal{T}(\tau)$ with
\begin{equation}\label{e:front1}
\mathcal{T} \equiv \frac{1}{\mathcal{R}}\left(
  \frac{\kappa^2\Sigma_0^\spost}{2\Omega\Sigma_c}\tau
  - 2 \Omega\int_0^\tau \mathcal{R} d\tau - \frac{\kxv^\spost}{\ky}\right),
\end{equation}
where $\mathcal{R}\equiv \Sigma_0^\spost/\Sigma_0=\uT/\uT^\spost$ is
the local expansion factor, and
\begin{equation}\label{e:front3}
 \frac{\kxv^\spost}{\ky} = \frac{\mu^{1/2}}{\cs} \left[
 \frac{\omgR}{\ky} - \vT(\xSH)  \right].
\end{equation}
Note that $\mathcal{T}$ defined in equation (\ref{e:front1}) is
identical to that given in equation (2.20) of \citet{bal88}.  It
governs the spatial (or temporal in the Lagrangian sense) behavior
of $\kxv$ of perturbations in shearing and expanding flows.

\begin{figure}
\epsscale{1.1}\plotone{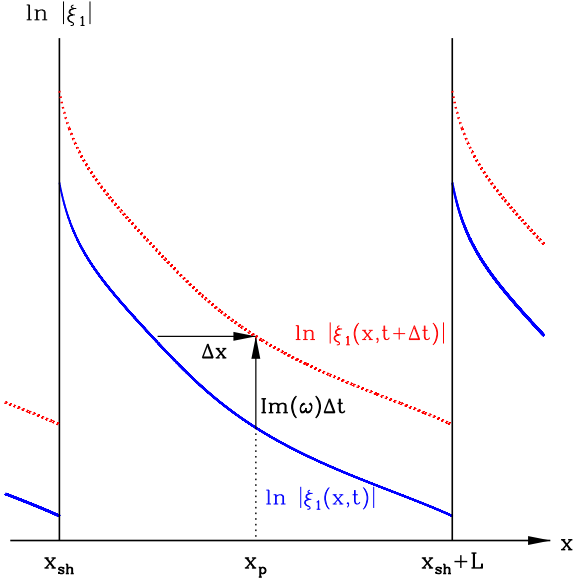} \caption{Schematic profile of the
perturbed PV for an unstable mode with $\omgI>0$ in between two
consecutive shocks located at $x=\xSH$ and $\xSH+L$. The solid and
dotted curves plot the amplitude of $\xi_1$ at time $t$ and
$t+\Delta t$, respectively.  The PV conservation requires that the
increased PV at $x=x_p$ during the time interval $\Delta t$ should
be equal to that advected from $x=x_p + \Delta x$, with $\Delta x =
-\uT \Delta t$.\label{f:pv_profile}}
\end{figure}

Equation (\ref{e:pvint1}) states that the amplitude of PV is
constant only for purely oscillatory modes with $\omgI=0$, whereas
it keeps decreasing (increasing) with $x$ away from the shock for
unstable (decaying) modes. Figure \ref{f:pv_profile} exemplifies the
situation with an unstable mode with $\omgI>0$, for which $\Xi_1$
grows in time. Since PV is preserved along a streamline, the
increased PV at a certain position $x=x_p$ during the time interval
$\Delta t$ due to instability should be equal to the advected PV
from the upstream position separated by $\Delta x = -\uT\Delta t$.
This is possible when $\Xi_1$ is a decreasing function of $x$ for
unstable modes.

In order for PV to be periodic in $x$, the spatial variation of
$\Xi_1$ inevitably requires a sudden change of the PV amplitude at
the shock front:
\begin{equation}\label{e:pvint3}
 \left| \frac{\Xi_1^\spost}{\Xi_1^\spre}\right| = e^{ 2\pi
 \omgI/\Omega},
\end{equation}
indicating that $|\Xi_1|$ should be enhanced (reduced) at the shock
for unstable (decaying) modes. The PV conservation in between spiral
shocks also requires that its phase should be different before and
after the shock. Since the elapsed time between two successive
shocks corresponds to $\tau=2\pi/\Omega$ and since
$\int_0^{2\pi/\Omega} \mathcal{R}d\tau = \Sigma_0^\spost L/(\Sigma_c
u_c) = 2\pi \Sigma_0^\spost/(\Sigma_c\Omega)$, equation
(\ref{e:front1}) demands that $\kxv/\Sigma_0$ should change at the
shock as
\begin{equation}\label{e:kjump}
 \frac{\kxv^\spre}{\Sigma_0^\spre} -
 \frac{\kxv^\spost}{\Sigma_0^\spost} = \frac{q\Omega L}{\uT\Sigma_0} \ky > 0.
\end{equation}
These changes in the amplitude and phase of the perturbed PV ought
to be consistent with the shock jump conditions that we derive in
the next subsection.

\subsection{Shock Jump Conditions}

\subsubsection{Perturbed Shock Front}

Perturbations of the form given in equation (\ref{e:ptb}), applied
to the background gas flow, also perturb the shock front into a
sinusoidal shape.  Define the shape of the perturbed shock front as
\begin{equation}\label{e:ps}
\zeta_1 (x,y,t) =  Z_1 \exp (-i\omega t + ik_yy),
\end{equation}
with amplitude $Z_1$.  Then,
\begin{equation}\label{e:sd}
s \equiv x - \xSH - \zeta_1
\end{equation}
measures the displacement from the moving shock front, with $s=0$
corresponding to the instantaneous shock location \citep{lee12}.

The unit vector normal to the instantaneous shock front is given by
\begin{equation}\label{e:nom}
\hat{\bf n} = \frac{\nabla s}{|\nabla s|} \simeq (1, -ik_y\zeta_1),
\end{equation}
to the first order in $\zeta_1$. On the other hand, the unit vector
tangent to the shock front is
\begin{equation}\label{e:tan}
\hat{\bf t} = \left(\left.\frac{dx}{dy}\right|_s, 1\right) =
(ik_y\zeta_1, 1),
\end{equation}
(see \citealt{dwa96,lee12}). Thus, the velocity of the shock front
is given as
\begin{equation}\label{e:vsh}
{\bf v}_{\rm sh} = \frac{\partial\zeta_1}{\partial t}\hat{\bf n} =
(-i\omega\zeta_1, 0),
\end{equation}
to the first order in $\zeta_1$.

The total gas surface density at the perturbed shock location can be
written as
\begin{equation}\label{e:ts}
\Sigma (\xSH + \zeta_1) = \Sigma_0(\xSH) + \Sigma_1 (\xSH) + \zeta_1
\left.\frac{d\Sigma_0}{dx}\right|_{\xSH},
\end{equation}
where the last term denotes the Taylor expansion of $\Sigma_0$ to
the perturbed shock position. The total gas velocity can similarly
be expanded to yield expressions for $\mathbf{v}_T (\xSH +
\zeta_1)$.

In the frame moving locally with the perturbed shock front, the
perpendicular component of the total velocity relative to the shock
front becomes
\begin{equation}\label{e:tpp}
\begin{split}
 v_\perp = & \mathbf{v}_{T0} (\xSH + \zeta_1) \cdot \hat{\bf n}
      - {\bf v}_{\rm sh} \cdot\hat{\bf n}\\
 = & \uT + u_1 +\zeta_1 \frac{d\uT}{dx}  + i\omega_D \zeta_1 ,
\end{split}
\end{equation}
while the parallel component is
\begin{equation}\label{e:tpr}
\begin{split}
  v_\parallel =& \mathbf{v}_{T0} (\xSH + \zeta_1) \cdot \hat{\bf t}
  - {\bf v}_{\rm sh} \cdot\hat{\bf t}\\
 = & \vT + v_1 + \zeta_1 \frac{d\vT}{dx} + ik_y\zeta_1\uT.
\end{split}
\end{equation}
Note that all quantities are evaluated at $x=\xSH$ in equations
(\ref{e:tpp}) and (\ref{e:tpr}).

\subsubsection{Jump Conditions}

Now we are ready to apply the Rankine-Hugoniot jump conditions
across the shock fronts:
\begin{mathletters}\label{e:rk}
\begin{eqnarray}
\jump{ v_\perp \Sigma }         & = & 0, \label{e:rk1} \\
\jump{(\cs^2 + v_\perp^2)\Sigma}  & = & 0, \label{e:rk2} \\
\jump{ v_\parallel  }       & = & 0, \label{e:rk3}
\end{eqnarray}
\end{mathletters}
where $\Delta_s(f)$ again denotes the difference of $f$ between the
immediate preshock and postshock regions.

Plugging equations (\ref{e:ts}), (\ref{e:tpp}), and (\ref{e:tpr})
into equations (\ref{e:rk}), one can show that the zeroth-order
terms are identical to equations (\ref{e:sj}). Taking the
first-order terms, equations (\ref{e:rk}) are simplified to
\begin{mathletters}\label{e:j123}
\begin{equation}\label{e:j1}
\Sigma_0\uT \jump{S_1} + \jump{\Sigma_0 U_1} + iZ_1\omega_D^s
\jump{\Sigma_0}=0,
\end{equation}
\begin{equation}\label{e:j2}
\left(\frac{\uT^2 + \cs^2}{2\uT}\right)\jump{S_1} + \jump{U_1} + Z_1
\Delta_s \left[
\left(\frac{\uT^2-\cs^2}{2\uT^2}\right)\frac{d\uT}{dx}\right] =0,
\end{equation}
\begin{equation}\label{e:j3}
\jump{V_1} - Z_1 \left(\frac{\kappa^2}{2\Omega}\frac{u_c}{\cs^2} -
i\ky\right) \jump{\uT} =0,
\end{equation}
\end{mathletters}
where $\omega_D^s=\omega_D(\xSH)$.

\subsection{Expansion near the Sonic Point}\label{s:sonic1}

Equations (\ref{e:ds1}) and (\ref{e:du1}) indicate that just like in
the background steady spiral shocks, there are certain conditions
that the perturbed quantities should obey at the sonic point to give
regular solutions for the perturbation variables.  To obtain these
conditions, we expand $S_1$, $U_1$, and $V_1$ near $x=\xSO$ as
\begin{mathletters}\label{e:exp}
\begin{eqnarray}
 S_1 & = & A_0 + A_1 \DX + \mathcal O(\DX^2), \label{e:exp1} \\
 U_1/(R\Omega) & = & B_0 + B_1 \DX + \mathcal O(\DX^2), \label{e:exp2} \\
 V_1/(R\Omega) & = & C_0 + C_1 \DX + \mathcal O(\DX^2), \label{e:exp3}
\end{eqnarray}
\end{mathletters}
with $A_{0,1}$, $B_{0,1}$, and $C_{0,1}$ being dimensionless
constants.

Plugging equations (\ref{e:so0}) and (\ref{e:exp}) into equation
(\ref{e:ds1}), one can show that the zeroth-order and first-order
terms in $\DX$, respectively, yield
\begin{equation}\label{e:C0}
C_0 = \frac{ia\omgSP A_0 - (i\omgSP-2\alpha_1)B_0}{(2+i \tky a)},
\end{equation}
and
\begin{equation}\label{e:texp1}
\begin{split}
 & a (2\alpha_1-i\omgSP)A_1  - (2\alpha_1-i\omgSP)B_1 \\
 & = (i\omgSP\alpha_1 - ia \tky\beta_{1m})A_0
  + (4\alpha_2 + i\tky\beta_{1m})B_0 \\
 &  - i\tky\alpha_1 C_0 - (2+i\tky a)C_1,
\end{split}
\end{equation}
where $\omgSP=\omega/\Omega -(0.5 - \xSO/R + \beta_0)\tky$,
$\tky=R\ky$, and $\beta_{1m}=\beta_1-1$. Equation (\ref{e:C0})
implies that $V_1$ at the sonic point cannot be taken arbitrarily:
it should depend on $S_1$ and $U_1$ for transonic solutions to
exist.

The zeroth-order terms of equation (\ref{e:du1}) are identical to
equation (\ref{e:C0}). Its first-order terms give
\begin{equation}\label{e:texp2}
\begin{split}
 & ia^2\omgSP A_1 + a (4\alpha_1 - i\omgSP )B_1 \\
 & =  i\tky a^2\beta_{1m}A_0 
 + (i\omgSP\alpha_1 - 4a\alpha_2 - i \tky a\beta_{1m})B_0 \\
  & + 2\alpha_1 C_0 + a(i\tky a+2)C_1.
\end{split}
\end{equation}
On the other hand, the zeroth-order terms of equation (\ref{e:dv1})
result in
\begin{equation}\label{e:C1}
C_1 =  -i\tky aA_0  - \frac{u_c/(R\Omega)}{a^2}B_0 +
\frac{i\omgSP}{a}C_0.
\end{equation}

Once $A_0$ and $B_0$ are known, therefore, $C_0$ and $C_1$ can be
calculated from equations (\ref{e:C0}) and (\ref{e:C1}),
respectively, and $A_1$ and $B_1$ by solving equations
(\ref{e:texp1}) and (\ref{e:texp2}) simultaneously.  This implies
that the solutions near the sonic point are completely specified by
two constants $A_0$ and $B_0$.

\subsection{Method of Computation}

Our problem involves four perturbation variables ($S_1, U_1, V_1,
Z_1$), one eigenvalue ($\omega$), three boundary conditions (eqs.
[\ref{e:j1}]--[\ref{e:j3}]) and one constraint at the sonic point
(eq.\ [\ref{e:C0}]). Since all the equations are linear, we may
arbitrarily take the amplitude of one variable at the sonic point.
Thus, the problem poses a well-defined eigenvalue problem, with four
unknowns and four constraints.

In practice, we fix ${\rm Re}(A_0)={\rm Im}(A_0)=1$ at the sonic
point, and choose two trial complex values for $\omega$ and $B_0$,
which give the values of $S_1$, $U_1$, and $V_1$ as well as their
derivatives at $x=\xSO$. We integrate equations
(\ref{e:ds1})--(\ref{e:dv1}) from the sonic point both in the
forward direction to $x=\xSH+L$ and in the backward direction to
$x=\xSH$, and then apply the periodic conditions for the
perturbation variables. At the shock front, equation (\ref{e:j3})
gives $Z_1$, which can be used to check the first boundary condition
(\ref{e:j1}). If equation (\ref{e:j1}) is not satisfied within a
tolerance, we return to the first step and repeat the calculation by
changing $B_0$.  After equation (\ref{e:j1}) is satisfied, we check
the second boundary condition (\ref{e:j2}). If equation (\ref{e:j2})
is not fulfilled, we again return to the first step to change
$\omega$, and continue the calculation iteratively until all the
perturbed shock jump conditions are satisfied.

\begin{deluxetable*}{ccccccccc}
 \tablecaption{Eigenfrequencies of One-dimensional Perturbations\label{t:eigen}}
 \tablehead { & \multicolumn{2}{c}{$\F=0.03$} &
              & \multicolumn{2}{c}{$\F=0.05$} &
              & \multicolumn{2}{c}{$\F=0.1$ } \\
 \cline{2-3}\cline{5-6} \cline{8-9}\\
  mode  & $\omgR/\Omega$& $\omgI/\Omega$ &&
          $\omgR/\Omega$& $\omgI/\Omega$ &&
          $\omgR/\Omega$& $\omgI/\Omega$}
 \startdata
 1 & $0.000$ & $-2.657\times10^{-1}$ && 0.000 &$-3.727\times10^{-1}$ && 0.354 & $-6.363\times10^{-1}$ \\
 2 & $0.628$ & $-9.230\times10^{-2}$ && 0.692 &$-1.782\times10^{-1}$ && 0.809 & $-3.897\times10^{-1}$ \\
 3 & $1.448$ & $+4.438\times10^{-4}$ && 1.496 &$+1.222\times10^{-3}$ && 1.614 & $-1.078\times10^{-2}$ \\
 4 & $2.573$ & $-6.212\times10^{-3}$ && 2.628 &$-1.380\times10^{-2}$ && 2.740 & $+7.693\times10^{-3}$ \\
 5 & $3.907$ & $-5.982\times10^{-3}$ && 4.023 &$-2.758\times10^{-2}$ && 4.316 & $-3.565\times10^{-2}$ \\
 6 & $5.349$ & $-1.120\times10^{-2}$ && 5.554 &$-2.543\times10^{-2}$ && 6.019 & $-5.747\times10^{-2}$ \\
 7 & $6.853$ & $-1.560\times10^{-2}$ && 7.144 &$-2.618\times10^{-2}$ && 7.846 & $-5.831\times10^{-2}$ \\
 8 & $8.391$ & $-1.935\times10^{-2}$ && 8.759 &$-3.077\times10^{-2}$ && 9.663 & $-8.771\times10^{-2}$  \\
 9 & $9.946$ & $-1.537\times10^{-2}$ &&10.389 &$-2.555\times10^{-2}$ &&11.509 & $-7.421\times10^{-2}$  \\
 10& $11.521$& $-1.941\times10^{-2}$ &&12.033 &$-2.250\times10^{-2}$ &&13.352 & $-8.511\times10^{-2}$  \\
\enddata
\end{deluxetable*}

\section{Dispersion Relation}\label{s:disp}

\subsection{One-dimensional Modes}

In this section, we apply the method described in Section
\ref{s:method} to the case of one-dimensional perturbations with
$k_y=0$. We find there are a pure decaying mode (with $\omgR=0$ and
$\omgI<0$) for small $\F$, a single overstable mode (with
$\omgR\neq0$ and $\omgI>0$), and many underdamping modes (with
$\omgI<0$). Equations (\ref{e:ds1})--(\ref{e:dv1}) and
(\ref{e:j1})--(\ref{e:j3}) guarantee that if $(S_1, U_1, V_1)$ is a
solution with eigenvalue $\omega$, then its complex conjugate is
also a solution with eigenvalue $-\omega^*$, provided $k_y=0$. This
indicates that eigenfrequencies exist always as a pair such that the
imaginary parts are the same, while the real parts differ only in
sign. We thus limit to the modes with $\omgR \geq 0$ in this
subsection.

\begin{figure}[!th]
\epsscale{1.2}\plotone{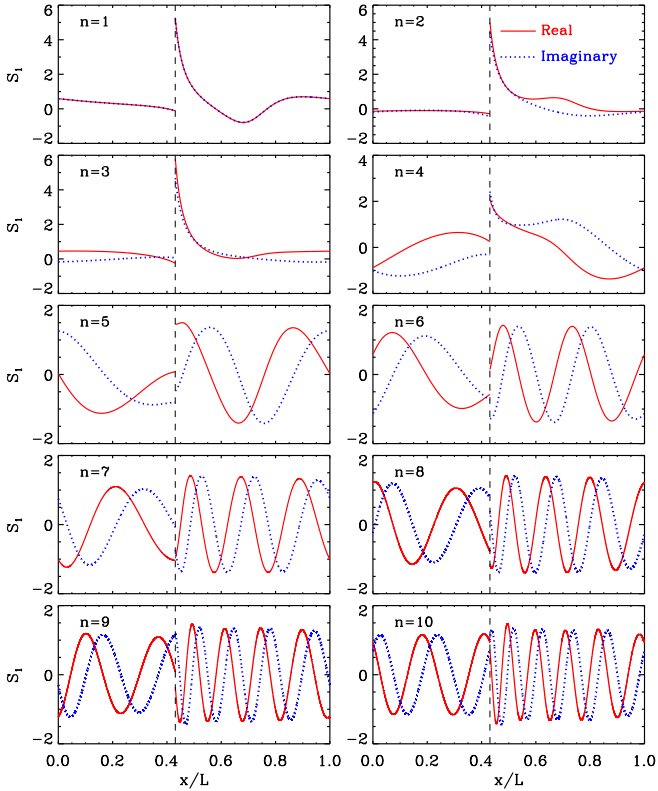} %
\caption{Ten lowest-frequency eigenfunctions $S_1$ of axisymmetric
modes with $\ky=0$ for $\F=5\%$. Solid and dotted lines correspond
to the real and imaginary parts of $S_1$, respectively, normalized
to unity at the sonic point located at $x/L=0.51$. They are
identical to each other for a pure decaying, $n=1$ mode. The
vertical line at $x/L=0.43$ in each panel marks the shock
front.\label{f:efun1d}}
\end{figure}

In Table \ref{t:eigen}, we list ten lowest eigenfrequencies for
$\F=3$, 5, and 10\%.  The modes are ordered in such a way that
$\omgRI{1}<\omgRI{2}<\omgRI{3}<\cdots$. Figure \ref{f:efun1d} plots
the corresponding profiles of the perturbed density $S_1$ for
$\F=5\%$, with the solid and dotted curves representing the real and
imaginary parts, respectively. Note that ${\rm Re}(S_1)= {\rm
Im}(S_1)$ for the $n=1$ mode since $\omgR=0$. The vertical dashed
line at $x/L=0.43$ marks the shock front. The number of nodes of the
eigenfunctions is $2(n-3)$ for modes with $n\geq 5$.  This indicates
that the wavenumber $\kx$ in the $x$-direction increases with
frequencies, which is a generic property of sound waves. We find
that the real parts of the eigenfrequencies are well fitted by
\begin{equation}\label{e:disp1d}
\omgR = (\langle \uT\rangle + \cs)\kx,
\end{equation}
with $\kx=2\pi(n-3)/L$ and the mean $x$-velocity $\langle
\uT\rangle/(L\Omega) = 0.175, 0.188$, and $0.216$ for $\F=3, 5$, and
$10\%$, respectively, which represents the ``spatially-averaged''
advection of sound waves by the background flow.

Sound waves propagating in a nonuniform medium naturally suffer
amplification or decay depending on the sign of the density and
velocity gradients relative to the propagation direction (e.g.,
\citealt{cla07}).  In addition, a shock wave not only reflects
incident sound waves but also amplifies them upon transmission
(e.g., \citealt{lan87,pij95}). In the case of galactic spiral
shocks, the eigenfrequencies given in Table \ref{t:eigen} show that
the non-uniform background and shock interactions usually dampen
sound waves, with a decay rate larger for larger $\F$. Note that
each spiral shock has a single mode with $\omgI>0$ ($n=3$ for $\F=3$
and $5\%$, and $n=4$ for $\F=10\%$).  This indicates that steady,
galactic spiral shocks are, in a strict sense, overstable to
one-dimensional displacements along the direction perpendicular to
the shock.  The overstable mode grows faster when the shock is
stronger and the background density varies more steeply. However,
the growth time of the overstability is
$\tgrow=2\pi/\omgI=8.2\times10^2/\Omega$ even for $\F=10\%$, which
is in general much longer than the Hubble time. This suggests that
these one-dimensional equilibrium spiral shocks can be regarded
stable for all practical purposes.

\subsection{Two-dimensional Modes}

We now search for two-dimensional normal modes with $\ky\neq0$ and
explore their stability.  For the numerical examples below, we focus
on the case with $\F=5\%$. The cases with different $\F$ are
qualitatively similar.

\subsubsection{Dispersion Relations}

\begin{figure}
\epsscale{1.2}\plotone{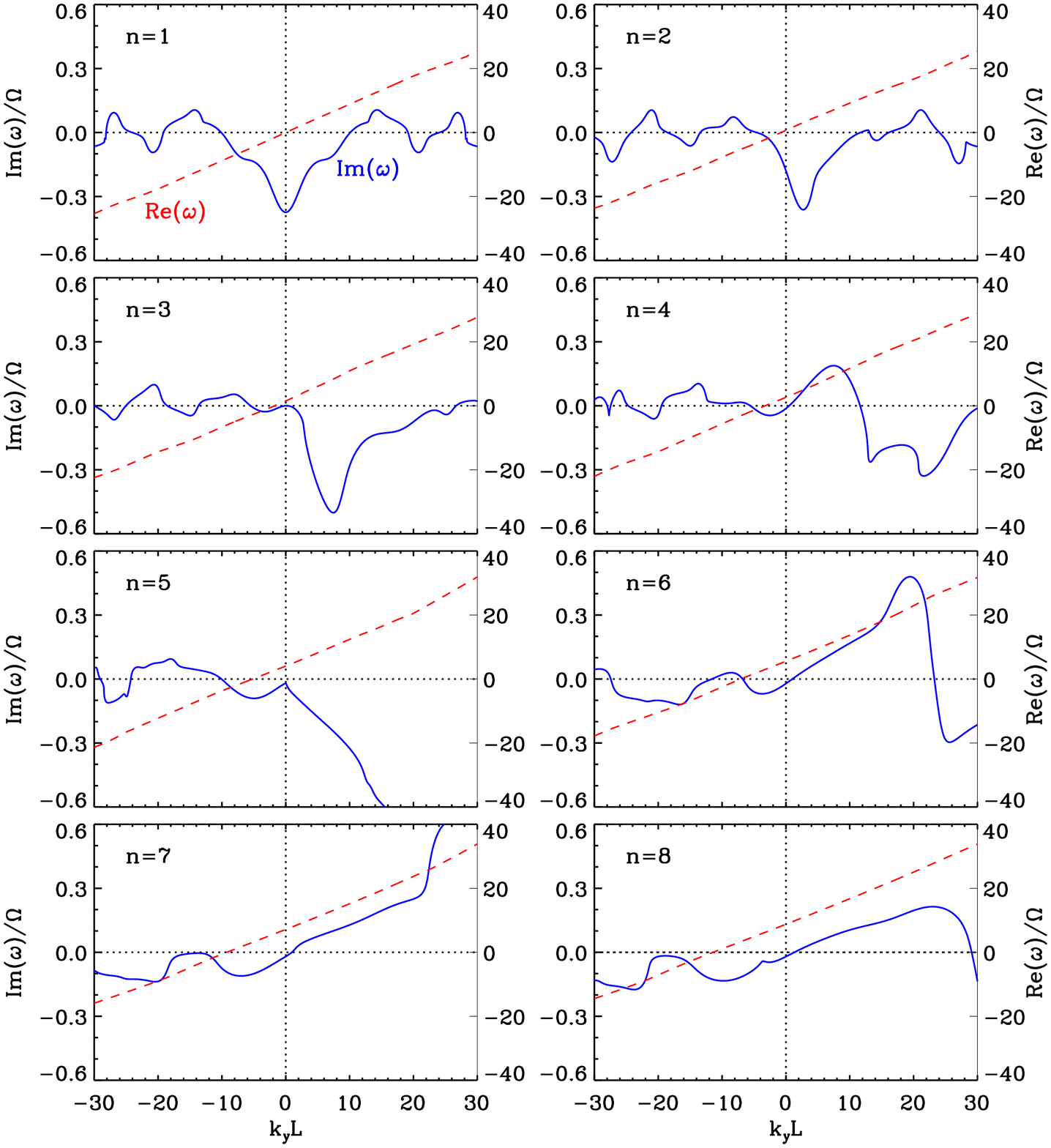} %
\caption{Non-axisymmetric dispersion relations of the eight
lowest-frequency eigenmodes for $\F=5\%$.  The modes are numbered in
the increasing order of $\omgR$ at $\ky=0$. In each panel, the blue
solid line (left $y$-axis) gives $\omgI$, while the red dashed line
(right $y$-axis) is for $\omgR$. The horizontal and vertical dotted
lines mark $\omega=0$ and $\ky=0$, respectively.\label{f:disp_all}}
\end{figure}

Unlike in the axisymmetric case, non-axisymmetric waves with
$\ky\neq0$ propagating in the positive and negative $y$-direction
behave differently from each other due to the non-vanishing $\vT$ in
the background flow, making $\omgR$ depend on the sign of $\ky$.
Figure \ref{f:disp_all} plots the dispersion relations of eight
lowest-frequency eigenmodes over $|\ky L|\leq 30$ for $\F=5\%$. As
in Figure \ref{f:efun1d}, these modes are numbered in the increasing
order of $\omgR$ at $\ky=0$. The dashed and solid lines show $\omgR$
and $\omgI$, respectively. It is apparent that $\omgR$ depends
almost linearly on $\ky$, with a slope of $\sim (0.80-0.86)
L\Omega$, indicating that the modes possess characteristics of
acoustic waves or entropy-vortex waves or their linear combinations.
Note that there are ranges of $\ky$ for which each mode becomes
overstable. However, the corresponding growth rate remains smaller
than $0.5\Omega$ except for the $n=7$ mode whose $\omgI$ keeps
increasing with $\ky$.

\begin{figure}
\epsscale{1.1}\plotone{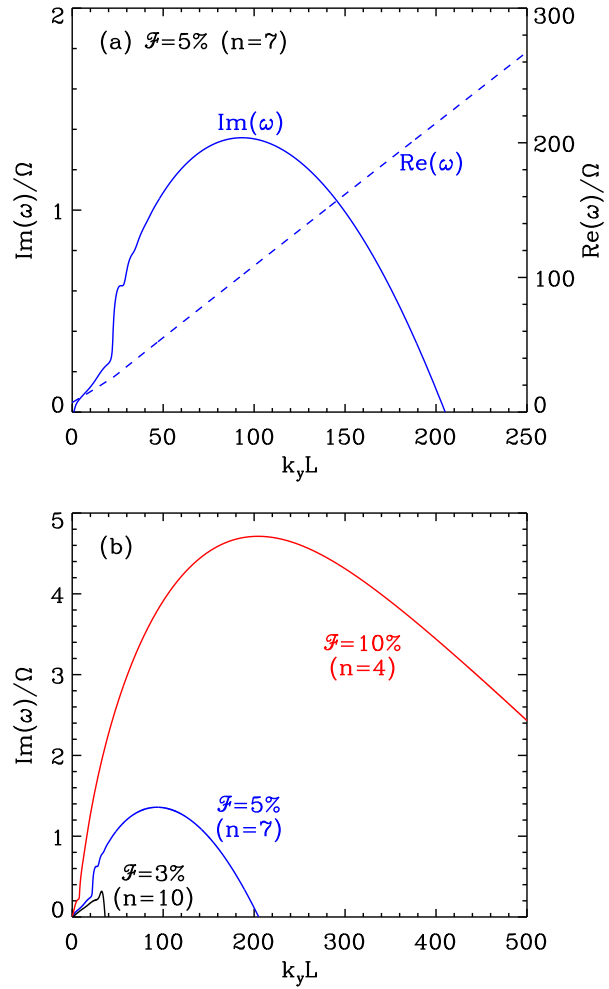} %
\caption{(a) Extended non-axisymmetric dispersion relation of the
$n=7$ mode for $\F=5\%$. The solid line (left $y$-axis) and the
dashed line (right $y$-axis) draw $\omgI$ and $\omgR$, respectively.
The maximum growth rate $\omgI_{\rm max}=1.36\Omega$ occurs at $\ky
L=92.4$. (b) Dependence of $\omgI$ on $\F$. The $n=10$, 7, and 4
modes are plotted for $\F=3$, 5, and 10\%, respectively.
\label{f:disp7}}
\end{figure}

Figure \ref{f:disp7}(a) plots an extended dispersion relation of the
$n=7$ mode for $\F=5\%$, while Figure \ref{f:disp7}(b) compares the
most unstable branches of the dispersion relations for differing
$\F$: the $n=10$, 7, and 4 modes are plotted, respectively, for
$\F=3$, 5, and 10\%. By searching for all possible overstable
modes, we have confirmed that these are the most unstable modes over
$|k_y L|\leq 500$ for given $\F$. The growth rate and wavelength of
the most unstable mode depend on $\F$ quite sensitively. The maximum
growth rate $\omgI_{\rm max}/\Omega=0.32$, 1.36, and 4.71 occurs at
$\ky L=32.2$, 92.4, and 204.5, with the corresponding real
eigenfrequency of $\omgR/\Omega = 38.1$, 100.6, and 198.3 for
$\F=3$, 5, and 10\%, respectively. Near the peak, $\omgI$ varies
slowly with $\ky$. When $\F=5\%$, for example, modes with $\ky L\sim
85$--$105$ have growth rates within 1\% of $\omgI_{\rm max}$. These
modes of WI would grow very rapidly and are likely to readily
manifest their presence in a galactic disk.

\begin{figure*}[!th]
\epsscale{0.8}\plotone{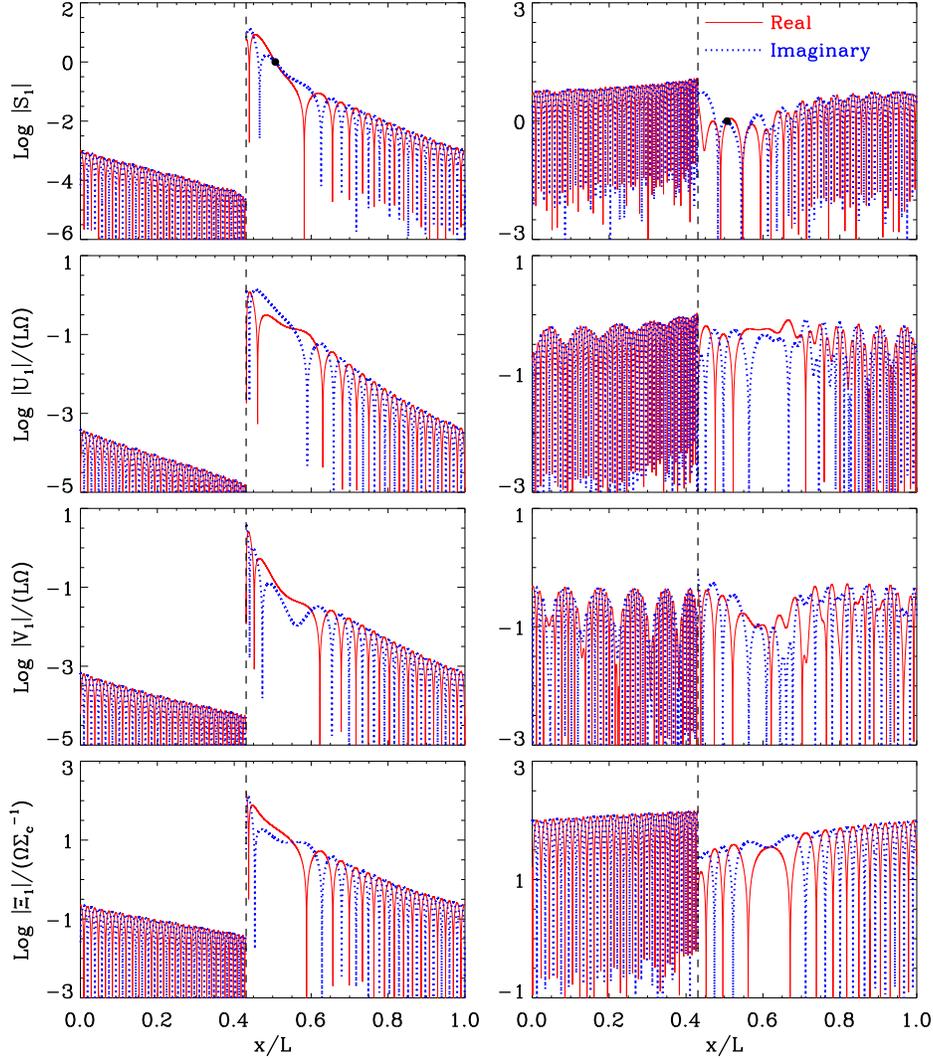} %
\caption{Eigenfunctions of (\emph{left column}) an unstable mode
with $\omega/\Omega = 100.6 + 1.36i$ and (\emph{right column}) a
decaying mode with $\omega/\Omega = 95.3 - 0.27i$. For both modes,
$\F=5\%$ and $\ky L=92.4$ are taken. The absolute values of the real
and imaginary parts of $S_1$, $U_1$, $V_1$, and $\Xi_1$ are plotted
as red solid and blue dotted curves, respectively. The vertical
dashed line in each panel marks the shock front, while black dots in
the top panels indicate the sonic point. \label{f:efun2d}}
\end{figure*}

The character of unstable or decaying modes can be identified by
exploring their eigenfunctions.  In the left panels of Figure
\ref{f:efun2d}, we plot the eigenfunctions $S_1$, $U_1$, $V_1$, and
$\Xi_1$ of the most unable mode with $\omega/\Omega = 100.6 + 1.36i$
for $\F=5\%$.  The decaying counterparts with $\omega/\Omega = 95.3
- 0.27i$ with the same $\ky$ and $\F$ are plotted in the right
panels, for comparison.  The vertical dashed line in each panel
marks the shock front, while the dots in the top panels indicate the
sonic point.  For the unstable mode, it is clear that the amplitudes
of the eigenfunctions decrease almost exponentially starting from
the shock front toward the downstream direction, and experience
large jumps at the shock, consistent with the prediction of equation
(\ref{e:pvint1}). For the decaying mode, on the other hand, the
amplitudes of the eigenfunctions except for $V_1$ increase with $x$
and exhibit sudden drops at the shock. For both cases, the
$x$-wavenumber $\kx$ of the perturbations increase as they propagate
away from the sonic point, making $\kx$ very large just before the
shock, while it has relatively small values in the postshock
regions. This spatial change of $\kx$ is due primarily to the
shearing and expanding background flow.

In general, any (linear) disturbance in the flow can be written as a
superposition of an entropy-vortex wave and an acoustic wave  (e.g.,
\citealt{lan87}):
\begin{equation}\label{e:decomp}
\begin{split}
S_1  &= S_{1,v} + S_{1,a}, \\
U_1  &= U_{1,v} + U_{1,a}, \\
V_1  &= V_{1,v} + V_{1,a},
\end{split}
\end{equation}
where the quantities with the subscripts ``$v$'' and ``$a$'' stand
for the contributions of the entropy-vortex and acoustic modes,
respectively. These waves would decouple from each other in a
uniform, non-rotating medium, but background gradients in the fluid
quantities as well as galactic rotation in galactic shocks tend to
mix them together unless their wavelengths are sufficiently small.
The eigenfunctions shown in Figure \ref{f:efun2d} suggest that the
WKB approximation (i.e., $\kx \gg |d\ln\uT/dx|$) is valid only in
the preshock regions. In this limit, one can show from equations
(\ref{e:con1})--(\ref{e:my1}) that the entropy-vortex modes with
$x$-wavenumber $\kxv$ are characterized by
\begin{mathletters}\label{e:ent0}
\begin{equation}\label{e:ent1}
 {\hat \omega_v}  \equiv \omgR - \uT \kxv - \vT \ky = 0,
\end{equation}
\begin{equation}\label{e:ent2}
 S_{1,v} = 2i\Omega/(\cs^2 \ky) U_{1,v},
\end{equation}
\begin{equation}\label{e:ent3}
 V_{1,v}  = - (\kxv/\ky) U_{1,v},
\end{equation}
\end{mathletters}
suggesting that these are incompressible and comoving with the
background flow. Note that equation (\ref{e:ent1}) is identical
to equation (\ref{e:pvint2}). For the acoustic modes, there are
various ways to construct a WKB dispersion relation, but the
acoustic parts of the decaying eigenfunctions presented in Figure
\ref{f:efun2d} turn out to be best described by
\begin{mathletters}\label{e:sonic0}
\begin{equation}\label{e:sonic1}
 {\hat \omega_a} \equiv \omgR - \uT \kxa - \vT \ky = -\cs \kxa,
\end{equation}
\begin{equation}\label{e:sonic2}
 S_{1,a} = (\kxa/\hat \omega_a) U_{1,a}=-U_{1,a}/\cs,
\end{equation}
\begin{equation}\label{e:sonic3}
 V_{1,a} = (\ky /\kxa ) U_{1,a},
\end{equation}
\end{mathletters}
which is free of vorticity, with $\kxa$ being the $x$-wavenumber of
acoustic waves. For $\kx \gg \ky$, entropy-vortex modes have
$|V_1|/|U_1| \gg 1$, while $\cs|S_1| \sim |U_{1}| \gg |V_1|$ for
acoustic modes, showing that most of the wave energy is contained in
$V_1$ for the former and in $S_1$ and $U_1$ for the latter.

\begin{figure*}[!th]
\epsscale{1.0}\plotone{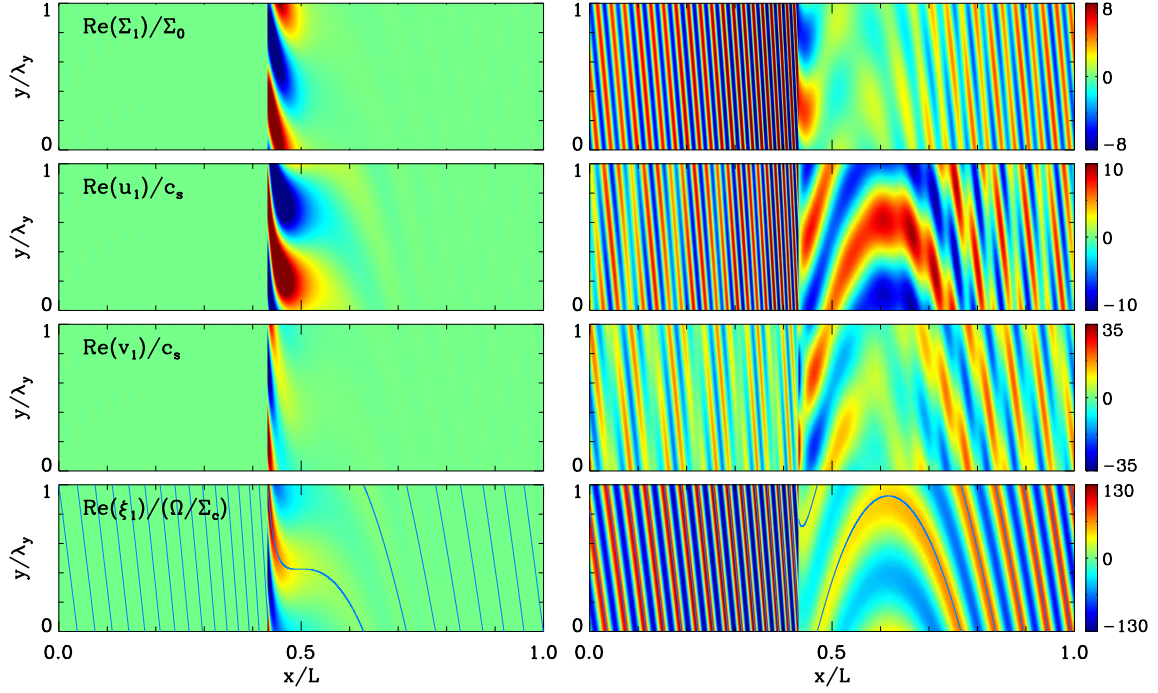} %
\caption{Distributions on the $x$-$y$ plane of the real parts of the
eigenfunctions of (\emph{left}) the unstable mode  and
(\emph{right}) the decaying mode shown in Figure \ref{f:efun2d}, for
Re$(\Sigma_1)/\Sigma_0$, Re$(u_1)/\cs$, Re$(v_1)/\cs$,
Re$(\xi_1)/(\Omega\Sigma_c^{-1})$ from top to bottom at $t=0$. The
ordinates are normalized by the wavelength $\lambda_y=2\pi/\ky =
0.068L$. The wavefronts of the perturbed PV are overlaid as solid
lines in the bottom panels. \label{f:tv_pv}}
\end{figure*}

Since $\kx$ becomes increasingly larger further downstream due
to shear, the decomposition of waves using equations
(\ref{e:decomp})--(\ref{e:sonic0}) is most meaningful in the regions
just before the shock front.  An inspection of the eigenfunctions
shown in the left panels of Figure \ref{f:efun2d} reveals that
$\kx^\spre L = 342.9$ and $V_1^\spre / U_1^\spre =-3.71$ in the
immediate preshock regions, which are almost equal to the
predictions of equations (\ref{e:ent1}) and (\ref{e:ent3}),
demonstrating that the unstable modes are predominantly an
entropy-vortex mode. On the other hand, the eigenfunctions of the
decaying modes have $\kxa^\spre L=1535$ and
$U_1^\spre/S_1^\spre=-0.97\cs$, roughly consistent with the
predictions of equations (\ref{e:sonic1}) and (\ref{e:sonic3}),
while $U_1^\spre/V_1^\spre \sim 3.2$ which cannot be described
solely by either equation (\ref{e:ent3}) or equation
(\ref{e:sonic3}). This indicates that both acoustic and
entropy-vortex modes contribute to the decaying mode, such that
$S_1$ and $U_1$ are dominated by the acoustic mode with
$\kxa^\spre/\ky=16.6$, while $V_1$ is affected by entropy-vortex
modes.  These results suggest that it is the entropy-vortex modes
that become unstable to the WI, while the acoustic modes play a
stabilizing role.

Using complex eigenfunctions, we can construct real perturbations as
\begin{equation}
\begin{split}
{\rm Re} (\Sigma_1)/\Sigma_0  =  e^{\omgI t}& \{ {\rm Re}(S_1)
\cos[\ky y-  \omgR t] \\& - {\rm Im}(S_1)\sin[\ky y- \omgR t]\},
\end{split}
\end{equation}
for the perturbed surface density, and similar expressions for the
other perturbation variables.  Figure \ref{f:tv_pv} plots real
eigenfunctions at $t=0$ on the $x$-$y$ plane of the (left column)
unstable and (right column) decaying mode shown in Figure
\ref{f:efun2d}. Note that the $y$-axis is normalized by the
perturbation wavelength $\lambda_y=2\pi/\ky$. The solid lines in the
bottom panels represent constant phases of PV obtained by
integrating equation (\ref{e:front1}) over $x$, which trace the
wavefront of the perturbed PV very well.  In the unstable case
shown, the perturbed density and velocity are dominated by the
entropy-vortex mode that is strongest in the postshock regions and
becomes weaker in the downstream direction (see eq.\
[\ref{e:pvint1}]). At the immediate behind of the shock front, they
have a trailing shape with $\kxv>0$, progressively rotate into a
less trailing shape due to shear reversal in the region with
$\Sigma_0/\Sigma_c\geq2$, and then become more trailing in the
interarm regions. For the decaying modes, on the other hand, the
wave amplitudes are stronger in the preshock regions, and the
$x$-wavenumber of $\Sigma_1$ and $u_1$ dominated by acoustic waves
is much larger than that of $v_1$ dominated by the entropy-vortex
waves.

\subsubsection{Physical Interpretation}

Since PV is preserved along a streamline in between shocks, the fact
that the WI relies on the entropy-vortex mode requires that
vorticity should be generated at the shock discontinuities.  In
Appendix \ref{a:vor}, we utilize the shock jump conditions
(\ref{e:j123}) to derive an expression for the PV changes at the
shock. In the WKB limit, equation (\ref{e:pvj1}) can be written as
\begin{equation}\label{e:pvj3}
\Xi_1^\spost - \Xi_1^\spre  \approx \Delta_{U_1} + \Delta_{Z_1} +
\Delta_{k_{x}},
\end{equation}
where
\begin{equation}\label{e:pvj4}
 \Delta_{U_1} \equiv  ik_y \frac{(\mu-1)^2}{\mu^2} \frac{U_1^\spre}{\Sigma_0^\spre},
\end{equation}
\begin{equation}\label{e:pvj5}
 \Delta_{Z_1} \equiv  -k_y \omega_D^s \frac{(\mu-1)^2}{\mu^2} \frac{Z_1}{\Sigma_0^\spre},
\end{equation}
and
\begin{equation}\label{e:pvj6}
 \Delta_{k_x} \equiv -i\ky \frac{q\Omega L }{\uT\Sigma_0}V_1^\spre
 = i \jump{\frac{\kxv}{\Sigma_0}}V_1^\spre.
\end{equation}
Note that $\Delta_{U_1}$ originates from the tangential variation of
the perpendicular velocity relative to the unperturbed shock, while
$\Delta_{Z_1}$ results from the deformation of a shock front itself
along the tangential direction. On the other hand, $\Delta_{k_x}$ is
due to the discontinuity of $\kx/\Sigma_0$ across the shock (eq.\
[\ref{e:kjump}]).

The role of the $\Delta$ terms in equation (\ref{e:pvj3}) in
producing or reducing PV differs from each other. The first
$\Delta_{U_1}$ term tends to decrease PV at the shock due to shock
compression of the perpendicular velocity. To see this more clearly,
for instance, let us consider a special case with $S_1=V_1 = Z_1=0$,
so that PV is contained in the $y$-variation of $U_1$. Then,
equations (\ref{e:pvb}) and (\ref{e:pva}) give
$\Xi_1^\spost/\Xi_1^\spre = (2\mu-1)/\mu^2 < 1$ for any $\mu\geq 1$,
showing that PV is reduced at the shock.  Also, the third
$\Delta_{k_x}$ term always tends to reduce PV across the shock,
which can be seen as follows. Since entropy-vortex modes usually
have $\kxv/\ky>1$ due to shear, $\Xi_1^\spre \approx i \omega_D^{sL}
V_1^\spre/(\uT\Sigma_0)$ from equation (\ref{e:pva}) is a good
approximation in the WKB limit. In this case, one can show that
$\Delta_{k_x} / \Xi_1^\spre \approx - (1+
\uT^\spre\kxv^\spre/[q\Omega L\ky])^{-1} < 0$. Physically, this
arises since the background shearing flow has
$\kxv^\spost/\kxv^\spre < \mu$ (e.g., equation [\ref{e:kjump}]),
which in turn makes $\Xi_1 \sim \kx V_1/\Sigma_0$ decreased after
the shock jump.

\begin{figure}
\epsscale{1.2}\plotone{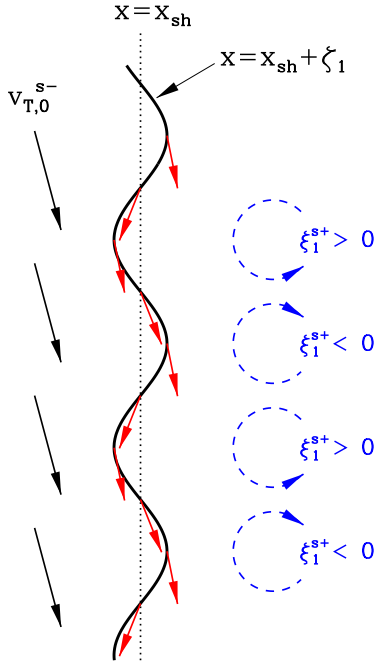} %
\caption{Schematic diagram showing the production of PV at the
distorted shock front ($x=\xSH + \zeta_1$). The vertical dotted line
indicates the unperturbed shock location ($x=\xSH$). The black
arrows on the left represent the background flow velocity
$\mathbf{v}_{T0}^\spre$ seen in the frame comoving with the deformed
shock front.  The red arrows indicate the directions of the flow
immediate after the shock. The resulting PV is positive at the
$y$-positions where $\zeta_1<0$ and negative where $\zeta_1>0$, as
indicated by the dashed arrows.\label{f:pv_sch}}
\end{figure}

On the other hand, the $\Delta_{Z_1}$ term is a \emph{source} for
the PV production at a deformed shock front, as Crocco's theorem
suggests. Figure \ref{f:pv_sch} schematically illustrates the PV
generation and the relationship between the signs of $\xi_1$ and
$\zeta_1$. The vertical dotted line and thick sinusoidal curve
indicate the unperturbed and perturbed shock fronts, respectively.
Since $\omgR/\ky > \vT(\xSH)$ (or $\omega_D^s >0$) from the linear
dispersion relation, the deformed shock front moves faster along the
$y$-direction than the background flow at the shock. Viewed in the
stationary shock frame, therefore, the background gas is moving in
the negative $y$-direction, as represented by the black arrows on
the left. In traversing the shock, the velocity vectors bend toward
the local tangent to an instantaneous shock front, which are
indicated by the red arrows. This naturally produces nonvanishing PV
(marked by dashed curved with arrows) in the postshock flow.  The
sign of the PV depends on the shape of the shock front, such that it
is positive (negative) in the regions where the shock is convex
(concave) seen from the upstream direction. That is, $\xi_1$ and
$\zeta_1$ have opposite signs, consistent with equation
(\ref{e:pvj5}).

When the $\Delta_{Z_1}$ terms dominates the other terms, PV
contained in the entropy-vortex waves can grow whenever the waves
pass through distorted spiral shocks in the course of galaxy
rotation. Interactions of traveling waves in the $x$-direction form
a standing entropy-vortex mode that can grow exponentially in time,
leading to the WI. That is, \emph{the WI refers to the growth of
entropy-vortex modes owing to vorticity generation from distorted
spiral shocks that the interstellar gas in galaxy rotation meets
periodically.} On the other hand, either when $U_1$ dominates the
perturbations, a most likely situation where acoustic modes are
stronger than entropy-vortex modes, or when $V_1$ dominates (without
involving strong shock deformations), PV drops at the shocks and the
associated entropy-vortex mode becomes weaker with time. This PV
reduction is responsible for the decaying modes shown in Figures
\ref{f:disp_all} and \ref{f:efun2d}. Since
$\Delta_{k_x}/\Delta_{Z_1} \propto \ky L/\omega_{D}^s$, the
$\Delta_{k_x}$ term becomes predominant for very large $\ky$,
eventually stabilizing the WI at $\ky L\simgt 205$ for the
dispersion relation displayed in Figure \ref{f:disp7}(a).

\section{Numerical Simulation}\label{s:num}

To check the wavelength and growth rate of the most unstable mode of
the WI found in the preceding section, we run direct numerical
simulations using the Athena code \citep{sto08,sto09}. Athena is an
Eulerian code for compressible magnetohydrodynamics based on
high-order Godunov schemes.  In this work, we use the constrained
corner transport method for directionally unsplit integration, the
HLLC Riemann solver for flux computation, and the piecewise linear
method for spatial reconstruction.

We first apply the Athena code to set up one-dimensional equilibrium
shock profiles for $\F=5\%$. The other galaxy and arm parameters are
taken the same as in the normal-mode analysis.  The simulation
domain has a length of $L$, which is resolved by 2048 zones. In
order to avoid strong non-steady gas motions induced by a sudden
introduction of the spiral potential, we increase its amplitude
slowly to make it achieve the full strength at $t\Omega=50$. The
system reaches a quasi-steady state at $t\Omega=100$, where the
density distribution consists of a steady part and a small-amplitude
fluctuating part. We have confirmed that the steady part is almost
identical to that shown in Figure \ref{f:1d}.

\begin{figure}
\epsscale{1.}\plotone{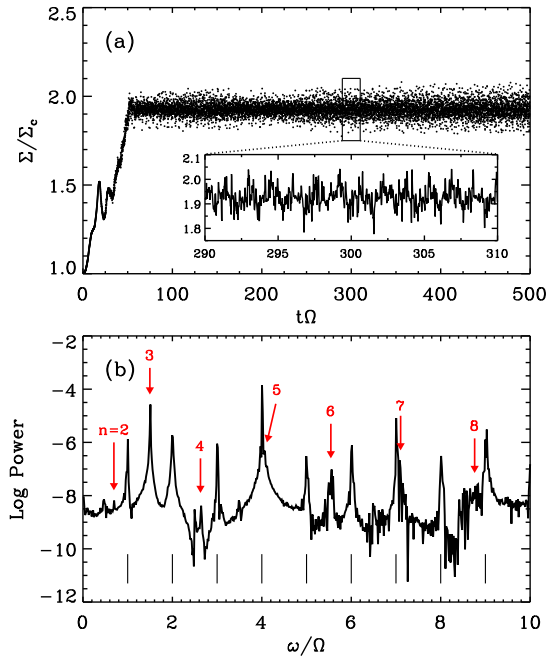} %
\caption{(a) Temporal variation of the gas surface density at
$x/L=0.5$ from a one-dimensional simulation with $\F=5\%$, and (b)
its power spectrum.  The inset in (a) zooms in the time range
$290\leq t\Omega \leq 310$ to clearly show fluctuations of
$\Sigma$. The presence of various modes makes $\Sigma$ fluctuate
with time, with its amplitude growing very slowly due to an
overstable ($n=3$) mode. The frequencies marked by short solid lines
in (b) correspond to the orbital crossing time and its higher
harmonics, while those indicated by red arrows represent the real
parts of the eigenvalues given in Table \ref{t:1d}. \label{f:1dsim}}
\end{figure}

\begin{figure*}
\epsscale{1.}\plotone{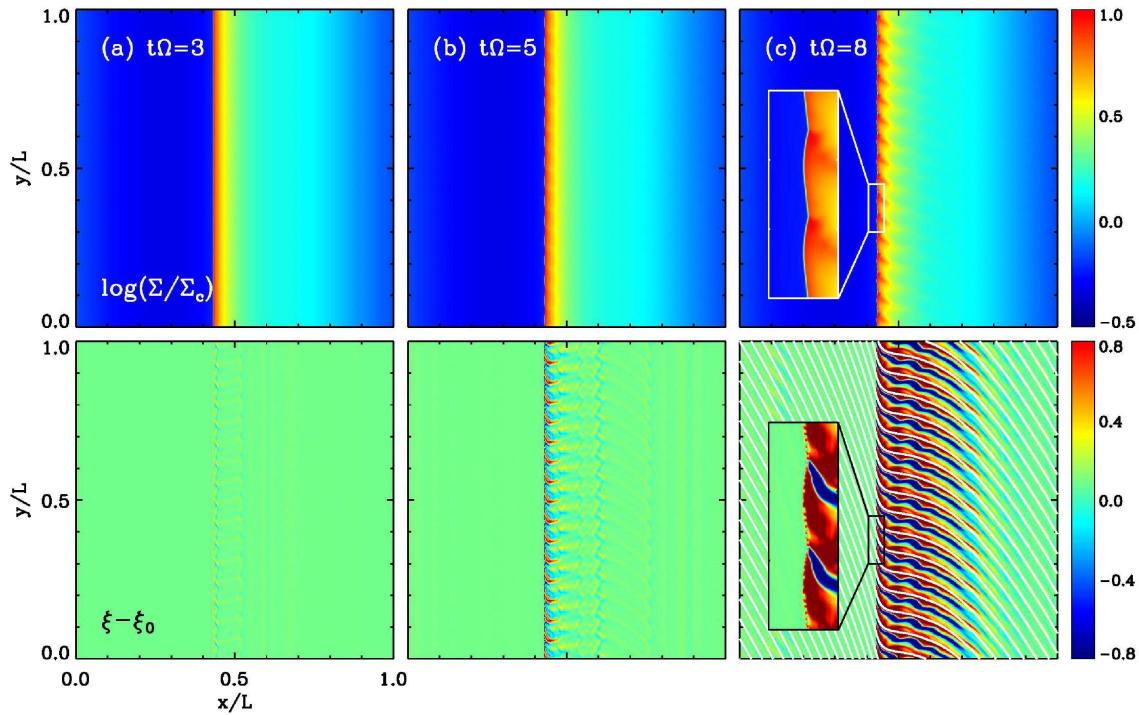} %
\caption{Snapshots of (upper panels) gas surface density and (lower
panels) perturbed PV at $t/\Omega=3$, 5, and 8 from a
two-dimensional run with $\F=5\%$ and $2048\times 2048$ resolution.
The wavefronts of the PV obtained by integrating equation
(\ref{e:front1}) are overlaid in the bottom right panel. The number
of the most unstable mode along the $y$-direction is 16 over the
distance of $L$.  The insets in (c) enlarge the section at $ 0.405
\leq x/L \leq 0.455$ and $ 0.30 \leq y/L \leq 0.45$. The upper and
lower colorbars label $\log(\Sigma/\Sigma_c)$ and $\xi - \xi_0$ in
units of $\Omega\Sigma_c^{-1}$, respectively. \label{f:2dmap}}
\end{figure*}

To examine the frequencies of the fluctuating density field, we
monitor the temporal evolution of the gas surface density at
$x/L=0.5$, which is plotted in Figure \ref{f:1dsim} together with
its Fourier-transformed power spectrum over $t\Omega = 200- 500$.
The mean and standard deviation of $\Sigma/\Sigma_c$ is $\sim 1.926$
and $0.040$, respectively. Note that the power spectrum is peaked at
some specific frequencies. The frequencies marked by the short line
segments at the bottom of Figure \ref{f:1dsim}(b) are the integral
multiples of $\Omega$, corresponding to the gas crossing time across
the simulation box and its higher harmonics. On the other hand, the
frequencies indicated by the arrows with numbers are very close to
those given in Table \ref{t:eigen}, indicating that these represent
decaying or growing eigenmodes identified in the normal-mode
analysis.  We note that among such modes, the $n=3$ mode with
$\omgR/\Omega=1.496$ has largest power since it is an overstable
mode.  Figure \ref{f:1dsim}(a) indeed shows a growing trend of the
gas surface density due to overstability, although the amplification
factor is only 60\% over $\Delta t\Omega=400$ because of too low a
growth rate.

Next, we simulate the WI of a spiral shock on the $x$-$y$ plane.  We
take the one-dimensional shock profile with $\F=5\%$ as a background
state.  We initially apply small-amplitude density perturbations
that are realized by a Gaussian random field with flat power, with a
standard deviation of $10^{-3}\Sigma_0$. For the simulation domain,
we set up a square box with size $L\times L$ and implement the
shearing box boundary conditions that can naturally handle shear in
the background flow \citep{haw95,kim02,kim06}. We set up a uniform
Cartesian grid with various resolutions.  Since the WI grows at
scales much smaller than $L$, it is necessary to run high-resolution
simulations to resolve it properly. We find that models with
$1024\times1024$ zones or higher give converged results, while those
with $512\times 512$ zones or less overestimate the wavelength of
the most unstable mode $\lambda_{y, \rm max}$. This suggests that
$\lambda_{y, \rm max}$ should be resolved by no smaller than 64
zones in order to accurately capture the WI.

\begin{figure}
\epsscale{1}\plotone{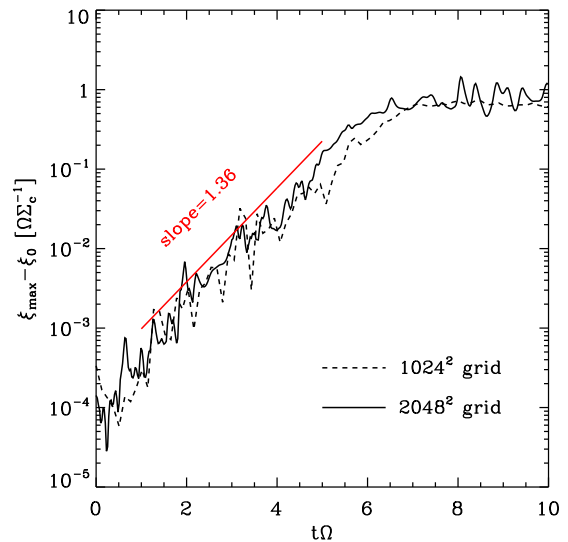} %
\caption{Evolution of the maximum PV measured at $x/L=0.45$ for the
$\F=5\%$ models with $1024\times1024$ and $2048\times2048$ zones.
The growth rates measured from the simulations are consistent with
the results of the normal-mode linear stability analysis, marked by
the line segment with slope of $1.36$. The WI saturates nonlinearly
at $t\Omega\simgt 7$. \label{f:2dgrow}}
\end{figure}

Figure \ref{f:2dmap} displays snapshots of (upper panels) density
structures in logarithmic scale and (lower panels) PV distributions
in linear scale at $t\Omega=3$, 5, and 8, from a run with
$2048\times2048$ resolution.  Figure \ref{f:2dgrow} compares the
time histories of the maximum PV relative to the initial value,
$\xi_{\rm max} - \xi_0$, measured at $x/L=0.45$ from models with
$1024\times1024$ and $2048\times2048$ zones. Initially, various
waves seeded by the density perturbations interact with the
background flow, and try to find eigenmodes that grow or decay
depending on the sign of $\omgI$.  The system soon picks up a few
modes that have large $\omgI$ and non-negligible initial power. At
$t\Omega=3$, the amplitudes of these modes are too small to be
readily discernible in the snapshots.  They keep growing during the
linear phase that lasts until $t\Omega\sim 7$, after which the
growth of the WI saturates. Figure \ref{f:2dmap} shows that the
wavenumber of the most strongly growing mode at $t\Omega \simgt 5$
is $\ky L\sim 32\pi = 100.5$, which is close to $k_{y, \rm
max}L=92.4$ predicted from the linear stability analysis. The
dominance of the $\ky L\sim100.5$ mode in the simulations is caused
by a combination of two facts: (1) its growth rate is only 0.5\%
smaller than, and is thus almost indistinguishable from that of the
most unstable mode (see Figure \ref{f:disp7}), and (2) it has an
initial amplitude about an order of magnitude larger than the latter
in our density perturbations.

The insets of Figure \ref{f:2dmap}(c) zoom in the section at $0.405
\leq x/L \leq 0.455$ and $0.30 \leq y/L \leq 0.45$ to clearly
display the distortion of the shock front. Note that $\xi -\xi_0$ is
positive (negative) in the regions where the shock is displaced
toward the upstream (downstream) direction, consistent with Figure
\ref{f:pv_sch}.  Consistent also with the shape of the
eigenfunctions, the density and PV distributions of the WI in the
numerical simulations have the shapes that are trailing as the gas
leaves the spiral shock, become less trailing in the region of shear
reversal $(0.43 \leq x/L \leq 0.50)$, and then become more trailing
afterwards. The growth rate of this mode measured from the
simulations is consistent with the prediction of the linear
stability analysis, plotted as a line segment with slope of $1.36$
in Figure \ref{f:2dgrow}.\footnote{We found that the background
shock in numerical simulations is not strictly stationary,
exhibiting small-amplitude motions in the $x$-direction due to the
presence of various modes with $\ky=0$ mentioned earlier. This
non-steady axisymmetric movement of the shock produces some spikes
in the evolutionary histories of $\xi_{\rm max}$ measured at a fixed
position.} The white lines overlaid in the bottom-right panel of
Figure \ref{f:2dmap} are the wavefronts obtained by integrating
equation (\ref{e:front1}), which are in good agreement with the PV
distributions in our simulations. All of these validate the results
of both our normal-mode stability analysis and numerical
simulations.

\section{Summary and Discussion}\label{s:sum}

We have presented the results of a normal-mode linear stability
analysis and hydrodynamic simulations for the WI of galactic spiral
shocks by employing a local shearing-box model of a galactic gaseous
disk under flat rotation.  We assume that the disk is
infinitesimally thin and remains isothermal, and do not consider the
effects of magnetic fields and gaseous self-gravity,  for
simplicity.  We first obtain one-dimensional profiles of
time-independent spiral shocks (Section \ref{s:back}). We then apply
small-amplitude perturbations to the steady solutions, and derive
the differential equations and the shock jump conditions that the
perturbation variables obey (Section \ref{s:method}). By solving the
perturbation equations as eigenvalue and boundary-value problems, we
obtain dispersion relations for various overstable and decaying
modes (Section \ref{s:disp}). We also compare the results of the
linear stability analysis with those of numerical simulations
(Section \ref{s:num}).

The dispersion relations show that there are various ranges of $\ky$
with which entropy-vortex modes become overstable, proving that the
WI is physical, rather than numerical, in origin. While PV remains
constant in a Lagrangian sense in between shocks, we show that it
experiences a sudden change across a shock primarily by the
following three processes: (1) tangential deformation of a shock
front, (2) tangential variation of the perturbed velocity
perpendicular to the unperturbed shock, and (3) discontinuity of
$\kx/\Sigma_0$ across the shock.  The first one increases the PV at
the shock, as a consequence of Crocco's theorem. On the other hand,
the last two processes tend to decrease PV through shock compression
and shear reversal across the shock, respectively. When the first
process dominates, PV of a gas element keeps increasing whenever it
passes through spiral shocks on its way of galaxy rotation. The
continuous increase of PV in a Lagrangian sense is realized in our
Eulerian stability analysis by standing entropy-vortex modes that
grow exponentially with time, leading to the WI. Sound waves and the
jumps in $\kx/\Sigma_0$ tend to suppress the WI, with the
stabilizing effect of the latter predominating for very short
wavelength perturbations. For $\F=5\%$, the most unstable modes are
found to have a growth rate comparable to the orbital angular
frequency $\Omega$ occurring at $\ky L\sim 10^2$, although these
become larger for higher $\F$. We confirm that the growth rate and
wavelength of the most unstable mode found in the linear stability
analysis are consistent with the results of direct numerical
simulations.

The assertion of \citet{wad04} that the WI was due to the KHI
was based on their result that a shear layer behind the shock has
the Richardson number $J<1/4$. As they noted, however, $J>1/4$ is
only a necessary condition for stability \citep{cha61}, so that
$J<1/4$ should not be interpreted as an instability criterion for
KHI.  On the other hand, we have shown in the present work that the
WI relies on the vorticity generation from a deformed shock front.
Although both WI and KHI involve vorticity, they differ in several
remarkable ways. First, the PV generation in the WI necessitates the
presence of a shock, which in turn requires non-vanishing
perpendicular velocity $\uT$ and density compression factor $\mu>1$,
while the KHI occurs when $\uT=0$ and $\mu=1$. Second, the WI is
global in the sense that it requires successive passages of a gas
flow across spiral shocks, which is attained by the periodic
boundary conditions in the current Eulerian analysis. On the other
hand, the KHI, when interpreted in terms of vorticity dynamics,
occurs as vorticity produced by disturbing an interface between two
fluids moving in opposite directions is accumulated at points where
it amplifies the interface distortion (e.g., \citealt{bat67,dra02}),
indicating that it is local.  Third, the WI is stabilized at very
large $\ky$ by shear reversal across the shock front, while the KHI
grows faster at larger $\ky$ (in the absence stabilizing agents such
as viscosity, conduction, etc., that are not considered in this
work). These differences clearly indicate that the WI studied
in this work cannot be attributed to KHI.  Referring to the work of
\citet{wad04}, \citet{ren13} mentioned KHI as a clump formation
mechanism in their high-resolution simulations. But, the locations
(postshock regions) and overall shapes (trailing as they leave the
shocks) as well as spacing ($\sim0.2\kpc$) of clumps shown in
their Figure 13 are similar to those of the eigenfunctions and wavelength
 of the WI (e.g., Figs.\ \ref{f:tv_pv} and \ref{f:2dmap}), suggesting
that they are the products of the WI rather than the KHI.

\citet{dob06} presented the results of SPH simulations for cloud
formation in spiral galaxies without considering the effects of
magnetic fields and self-gravity (see also \citealt{dob_etal06}).
They showed that spiral shocks efficiently form gas clumps that are
sheared out in the interarm regions to appear as feathers only if
gas is cold, while warm gas with $T>10^4$ K is unable to produce
clumps. They interpreted the clump formation in their cold-gas model
as being arising from angular momentum exchanges among particles in
the shock that are already inhomogeneous before entering the shock.
Similarities between their density maps and those in \citet{wad04}
strongly suggest that the clump formation in their SPH models is
most likely due to the WI. In the picture of WI, colder gas is more
prone to the instability since it induces stronger shocks,
corresponding to larger $\F$.

In this work we take an eigenvalue approach to analyze the stability
of spiral shocks. \citet{dwa96} took another approach by solving a
linearized set of hydrodynamic equations as an initial value problem
subject to the shock jump conditions.  Since the two methods are
complementary to one another, they should yield the same results.
However, \citet{dwa96} reported that non-self-gravitating and
unmagnetized shocks are linearly stable due possibly to the
non-vanishing radial velocity, which is seemingly in contrast to our
results.  We note that their conclusion was based on long-wavelength
perturbations with $\ky L\sim 1$, about two orders of magnitude
smaller than $k_{y, \rm max}$ found in the present work, which were
evolved only up to $t\Omega\sim1$. Figure \ref{f:disp_all} shows
that modes with $|\ky| L\le 1$ can also be unstable, although their
growth rates are less than $\sim0.015\Omega^{-1}$.  Since the
corresponding amplification factor over one orbital period is less
than 10\%, they were unlikely to grow to appreciable amplitudes in
the work of \citet{dwa96}. In addition, such slowly growing modes
could easily be suppressed by numerical viscosity present in any
numerical scheme (e.g., \citealt{kimcg10,kim_sto12}).

While we have shown that the WI grows very rapidly in a razor-thin
disk with no magnetic field, it still remains to be seen whether it
is responsible for dense arm clouds and feathers in real spiral
galaxies for the following two reasons. First, it is unclear whether
the WI would operate in disks that are magnetized and vertically
stratified. \citet{kim06} showed that spiral shocks exhibit flapping
motions in the direction perpendicular to the arm, when the vertical
degree of freedom is considered.  These motions are caused by
incommensurability between the arm-to-arm crossing time with the
vertical oscillation periods, capable of injecting turbulent energy
into dense post-shock gas \citep{kimcg06,kimcg08}. These non-steady
motions as well as strong vertical shear present in
three-dimensional shocks appear to disrupt coherence of vortical
structures at different heights, preventing the growth of WI
\citep{kim06}.  In addition, the presence of magnetic fields appears
to stabilize WI \citep{dob_price08} and completely quenches it when
the fields are of equipartition strength or stronger \citep{she06},
although it is uncertain whether the magnetic stabilization is due
to magnetic forces on the perturbations or through a reduced
background shock strength.

Second, even if the WI does develop in real disk galaxies, the
connection between the WI and observed giant clouds and interarm
feathers is not direct.  The WI itself involves perturbations only
near the shock front, resulting in very weak perturbations in the
interarm regions. It also occurs at very small spatial scales,
corresponding to 0.07 times the arm-to-arm spacing when $\F=5\%$ and
even smaller scales when the arms are stronger.  On the other hand,
observed feathers and giant clouds in the arms of M51 have a mean
separation of order $\sim(0.5-2)$ kpc (e.g., \citealt{elm83,sch13}),
consistent with the Jeans length at the arm density peak (e.g.,
\citealt{elm94,kim02}), and appear quite strong also in the interarm
regions. Recently, \citet{lee12} carried out a linear stability
analysis of feathering instability by including self-gravity and
magnetic fields. They showed that feathering modes retain relatively
strong presence in the interarm regions and grow sufficiently
rapidly, indicating that self-gravity may be essential for the
formation of interarm feathers.  All of these suggest that the
WI \emph{alone} is unlikely responsible for interarm feathers. Of
course, it cannot be ruled out the possibility that small clumps
produced primarily by the WI become denser by self-gravity,
radiative cooling, and/or through mutual mergers, as in
high-density clumps produced in models of \citet{ren13}, possibly
developing into feathers in the downstream side. It would thus be
interesting to explore the effects of self-gravity and magnetic
fields on the WI, and their relationships with nonaxisymmetric
interarm features.

\acknowledgments

We gratefully acknowledge constructive comments from the referee, as
well as helpful discussions with B.~G.~Elmegreen and E.~C.~Ostriker.
This work was supported by the National Research Foundation of Korea
(NRF) grant funded by the Korean government (MEST), No.\
2010-0000712.

\appendix

\section{Jump of Potential Vorticity at the Perturbed Shock Front}\label{a:vor}

Galactic gas flows periodically meet spiral shocks, once in every
$2\pi/\Omega$ interval. While PV is conserved along a given
streamline in between shocks, it inevitably experiences a sudden
jump when moving across a distorted shock. In this Appendix, we
derive the jump condition for the perturbed PV, $\jump{\Xi_1} =
\Xi_1^\spost - \Xi_1^\spre$, at the shock front ($x=\xSH$).

We first want to express $S_1^\spost$, $U_1^\spost$, and
$V_1^\spost$ in terms of $S_1^\spre$, $U_1^\spre$, $V_1^\spre$, and
$Z_1$. It is useful to write
\begin{equation}\label{e:aux2}
 \uT^\spost= \cs \mu^{-1/2},\;\;\;\text{and}\;\;\;  \uT^\spre=
 \cs\mu^{1/2},
\end{equation}
from equations (\ref{e:bd1}) and (\ref{e:mu}). It then follows that
\begin{equation}\label{e:app1}
 \frac{d}{dx} \ln (\uT^\spost\uT^\spre)  =
 -(\mu-1) \frac{d\ln\uT^\spre}{dx},
\end{equation}
from equation (\ref{e:momx01}).

With the help of equations (\ref{e:aux2}) and (\ref{e:app1}), we
solve equations (\ref{e:j1}) and (\ref{e:j2}) for $S_1^\spost$ and
$U_1^\spost$ to obtain
\begin{equation}\label{e:pjs}
  S_1^\spost = S_1^\spre +
  \frac{2}{\cs\mu^{1/2}} \left( U_1^\spre
  + i\omega_D^s Z_1
  - \frac{\mu-1}{2} \frac{d\uT^\spre}{dx} Z_1 \right),
\end{equation}
\begin{equation}\label{e:pju}
  U_1^\spost = - \frac{1}{\mu} U_1^\spre
  -i \omega_D^s \left(1+\frac{1}{\mu} \right) Z_1
  + \frac{\mu-1}{\mu} \frac{d\uT^\spre}{dx} Z_1,
\end{equation}
for $\mu\neq1$.\footnote{For $\mu=1$, equations (\ref{e:j1}) and
(\ref{e:j2}) yield of course a trivial solution $(S_1^\spost,
U_1^\spost) = (S_1^\spre, U_1^\spre)$.} Here, $\omega_D^s =
\omega_D(\xSH)$. Equation (\ref{e:j3}) simply results in
\begin{equation}\label{e:pjv}
V_1^\spost = V_1^\spre -
\frac{\mu-1}{\mu^{1/2}}\left(\frac{\kappa^2}{2\Omega}\frac{u_c}{\cs}
-ik_y\cs\right) Z_1.
\end{equation}
These are the jump conditions that the perturbation variables should
satisfy at the shock front.

Substituting equation (\ref{e:dv1}) in equation (\ref{e:pv1}) and
arranging the terms using equations (\ref{e:pjs})--(\ref{e:pjv}), we
obtain the perturbed PV immediate behind of the shock
\begin{equation}\label{e:pvb}
\begin{split}
\Xi_1^\spost = & \left( ik_y \frac{1-2\mu}{\mu^2} -
\frac{\kappa^2}{2\Omega} \frac{u_c}{\mu\cs^2} \right)
 \frac{U_1^\spre}{\Sigma_0^\spre}
 - \left( ik_y \frac{\cs^2}{\uT^\spre} +  \frac{\kappa^2}{2\Omega} \frac{u_c}{\uT^\spre}
 \right)
 \frac{S_1^\spre}{\Sigma_0^\spre} \\
& + \frac{i \omega_D^s}{\uT^\spre} \frac{V_1^\spre}{\Sigma_0^\spre}
  - k_y \omega_D^s \frac{(\mu-1)^2}{\mu^2} \frac{Z_1}{\Sigma_0^\spre}
  + ik_y  \frac{(\mu-1)^2}{\mu^2} \frac{d\uT^\spre}{dx}
  \frac{Z_1}{\Sigma_0^\spre},
\end{split}
\end{equation}
while the perturbed PV ahead of the shock is given by
\begin{equation}\label{e:pva}
\begin{split}
\Xi_1^\spre = &
 -\left( ik_y  + \frac{\kappa^2}{2\Omega} \frac{u_c}{\mu\cs^2}
 \right)
 \frac{U_1^\spre}{\Sigma_0^\spre}
 - \left( ik_y \frac{\cs^2}{\uT^\spre} +  \frac{\kappa^2}{2\Omega} \frac{u_c}{\uT^\spre}
 \right)
 \frac{S_1^\spre}{\Sigma_0^\spre} +
 \frac{i \omega_D^{sL}}{\uT^\spre} \frac{V_1^\spre}{\Sigma_0^\spre},
\end{split}
\end{equation}
where $\omega_D^{sL} = \omega_D(\xSH+L) = \omega_D^s + q\Omega
L\ky$.

Subtraction of equation (\ref{e:pva}) from equation (\ref{e:pvb})
gives the jump condition for the PV across the perturbed shock
\begin{equation}\label{e:pvj1}
\jump{\Xi_1}  =
  ik_y \frac{(\mu-1)^2}{\mu^2} \frac{1}{\Sigma_0^\spre} \left( U_1^\spre   +
   \frac{d\uT^\spre}{dx}
  Z_1   + i\omega_D^s
  Z_1 \right)-  i \ky \frac{q\Omega L}{\uT\Sigma_0}V_1^\spre,
\end{equation}
which can be rewritten in a more illuminating form as
\begin{equation}\label{e:pvj2}
  \jump{\xi_1} = \frac{(\mu-1)^2}{\mu^2}  \frac{1}{\Sigma_0^\spre}
  \frac{\partial v_{\perp, 1}^\spre} {\partial y} -  \frac{q\Omega
  L}{\uT\Sigma_0} \frac{\partial v_1^\spre}{\partial x},
\end{equation}
with
\begin{equation}
v_{\perp, 1}^\spre=  u_1^\spre +  \zeta_1(d\uT^\spre/dx)
  + i\omega_D^s \zeta_1,
\end{equation}
being the perturbed preshock velocity perpendicular to the
instantaneous shock front (see eq.\ [\ref{e:tpp}]). The second term
in the right-hand side of equation (\ref{e:pvj2}) follows from
equations (\ref{e:pv1}) and (\ref{e:kjump}), resulting originally
from non-uniform background shear across a shock front.  Equation
(\ref{e:pvj2}) states that the PV jump at the shock is due to two
factors : the tangential variation of the perpendicular velocity and
the discontinuous change of $\kxv/\Sigma_0$ at the shock.

The origin of the first term in equation (\ref{e:pvj2}) is Crocco's
theorem for vorticity generation from a curved shock front.
\citet{hay57} showed that the vorticity jump across an unsteady
shock amounts to
\begin{equation}\label{e:crocco}
\jump{\nabla \times \mathbf{v}|_z}  = -\frac{(\mu-1)^2}{\mu}
\left(V_s K + \frac{\partial C_r}{\partial S}\right),
\end{equation}
where $K$ is the curvature of the shock front, $V_s$ is the
tangential component of the fluid velocity, $C_r$ is the shock speed
relative to the normal component of the preshock fluid velocity, and
$S$ denotes the coordinate tangential to the shock (see also
\citealt{tru52,kev97}). Noting that spiral shocks in our local
models are straight $(K=0)$, one can see that $\jump{\Xi_1}=
\jump{\nabla \times \mathbf{v}|_z} /\Sigma_0^\spre$ with
$C_r=-v_{\perp, 1}^\spre$ and $S=y$ in the absence of background
shear ($q\Omega=0$).

\begingroup
\sloppy

\endgroup

\end{document}